\documentclass[12pt,a4paper]{article}
\usepackage[usenames,dvipsnames]{xcolor}
\usepackage{jheppub}  
\usepackage{amssymb} 
\usepackage{amsmath}
\usepackage{amsthm}
\usepackage{mathtools}
\usepackage{amsfonts}    
\usepackage{dsfont}
\usepackage{pdfpages}
\usepackage{verbatim}
\usepackage{tensor}
\usepackage{mathrsfs}
\usepackage{textgreek}
\usepackage{braket}
\usepackage{tikz}
\usetikzlibrary{decorations.pathreplacing, decorations.markings,calc,shapes.misc,decorations.pathmorphing,patterns.meta, math}
\tikzset{snake it/.style={decorate, decoration=snake}}
\tikzset{cross/.style={cross out, draw=black, minimum size=2*(#1-\pgflinewidth), inner sep=0pt, outer sep=0pt},
cross/.default={1pt}}
\usetikzlibrary{shapes.geometric}
\tikzset{
    partial ellipse/.style args={#1:#2:#3}{
        insert path={+ (#1:#3) arc (#1:#2:#3)}
    }
}
\usepackage{listings}
\usepackage[most]{tcolorbox}

\definecolor{codegreen}{rgb}{0,0.6,0}
\definecolor{codegray}{rgb}{0.5,0.5,0.5}
\definecolor{codepurple}{rgb}{0.58,0,0.82}
\definecolor{backcolour}{rgb}{0.95,0.95,0.92}

\definecolor{awesome}{rgb}{1.0, 0.13, 0.32}

\lstdefinestyle{mystyle}{
    backgroundcolor=\color{backcolour},   
    commentstyle=\color{codegreen},
    keywordstyle=\color{magenta},
    numberstyle=\tiny\color{codegray},
    stringstyle=\color{codepurple},
    basicstyle=\ttfamily\footnotesize,
    breakatwhitespace=false,         
    breaklines=true,                 
    captionpos=b,                    
    keepspaces=true,                 
    numbers=left,                    
    numbersep=5pt,                  
    showspaces=false,                
    showstringspaces=false,
    showtabs=false,                  
    tabsize=2
}

\hypersetup{
    pdfencoding=unicode,
	colorlinks=true,
	urlcolor=ForestGreen,
	linkcolor=ForestGreen,
	citecolor=ForestGreen,
	pdftitle={HEM and ground ring},
	pdfauthor={Aleksandr Artemev},
	pdfdisplaydoctitle=true,
	pdfstartview=FitH,
	linktocpage=true
}

\newcommand{\ba}{\begin{align}}

\newcommand{\be}{\begin{equation}}
\newcommand{\ee}{\end{equation}}
\def\bd{\begin{tikzpicture}}
\def\ed{\end{tikzpicture}}

\newcommand\Res{\mathop{\text{Res}}}
\allowdisplaybreaks[1]

\newcommand\ZZ{\mathbb{Z}}

\newcommand\NN{\mathbb{N}}

\DeclareMathOperator\Disc{Disc}

\renewcommand\d{\text{d}}
\newcommand{\normord}[1]{:\mathrel{#1}:}

\usepackage{amsmath,tikz}
\usepackage{subcaption}
\usetikzlibrary{patterns,arrows.meta}
\definecolor{bleudefrance}{rgb}{0.19, 0.55, 0.91}

\definecolor{vert}{rgb}{0.1367 0.543 0.1367}

\tikzset{
  pics/diske/.style n args={1}{
    code = { %
        \filldraw[color=black, fill=black!30, thick] circle (0.3);
        \node[color=black] at (0,0.5) {\scriptsize ${#1}$};
    }
  }
}
\tikzset{
  pics/disker/.style n args={1}{
    code = { %
        \filldraw[color=red, fill=black!30, thick] circle (0.3);
        \node[color=black] at (0,0.5) {\scriptsize ${#1}$};
    }
  }
}
\tikzset{
  pics/diskeb/.style n args={1}{
    code = { %
        \filldraw[color=blue, fill=black!30, thick] circle (0.3);
        \node[color=black] at (0,0.5) {\scriptsize ${#1}$};
    }
  }
}   
\tikzset{
  pics/disk1/.style n args={1}{
    code = { %
        \filldraw[color=black, fill=black!30, thick] circle (0.3);
        \node[color=black] at (0,0.5) {\scriptsize ${#1}$};
        \draw node[cross=3pt, very thick] {};
    }
  }
}   
\tikzset{
  pics/disk1r/.style n args={1}{
    code = { %
        \filldraw[color=red, fill=black!30, thick] circle (0.3);
        \node[color=black] at (0,0.5) {\scriptsize ${#1}$};
        \draw node[cross=3pt, very thick] {};
    }
  }
}
\tikzset{
  pics/disk1b/.style n args={1}{
    code = { %
        \filldraw[color=blue, fill=black!30, thick] circle (0.3);
        \node[color=black] at (0,0.5) {\scriptsize ${#1}$};
        \draw node[cross=3pt, very thick] {};
    }
  }
}
\tikzset{
  pics/disk2/.style n args={1}{
    code = { %
        \filldraw[color=black, fill=black!30, thick] circle (0.3);
        \node[color=black] at (0,0.5) {\scriptsize ${#1}$};
        \node[cross=3pt, very thick] at (0.125,0) {};
        \node[cross=3pt, very thick] at (-0.125,0) {};
    }
  }
}
\tikzset{
  pics/disk2r/.style n args={1}{
    code = { %
        \filldraw[color=red, fill=black!30, thick] circle (0.3);
        \node[color=black] at (0,0.5) {\scriptsize ${#1}$};
        \node[cross=3pt, very thick] at (0.125,0) {};
        \node[cross=3pt, very thick] at (-0.125,0) {};
    }
  }
}
\tikzset{
  pics/disk2b/.style n args={1}{
    code = { %
        \filldraw[color=blue, fill=black!30, thick] circle (0.3);
        \node[color=black] at (0,0.5) {\scriptsize ${#1}$};
        \node[cross=3pt, very thick] at (0.125,0) {};
        \node[cross=3pt, very thick] at (-0.125,0) {};
    }
  }
}

\tikzset{
  pics/disk1b1/.style n args={1}{
    code = { %
        \filldraw[color=black, fill=black!30, thick] circle (0.3);
        \node[color=black] at (0,0.5) {\scriptsize ${#1}$};
        \node[cross=3pt, very thick] at (0,0) {};
        \node[cross=3pt, very thick] at (.3,0) {};
    }
  }
}
\tikzset{
  pics/cylnn0/.style n args={2}{
    code = { %
        \filldraw[color=black,fill=black!30, thick] circle (0.35);
        \filldraw[color=black, fill=white, thick]  circle (0.17);
        \node[color=black] at (0,0.51) {\scriptsize ${#1}$};
        \node[color=black] at (0,0) {\scriptsize ${#2}$};
    }
  }
}
\tikzset{
  pics/cylnnrb0/.style n args={2}{
    code = { %
        \filldraw[color=red,fill=black!30, thick] circle (0.35);
        \filldraw[color=blue, fill=white, thick]  circle (0.17);
        \node[color=black] at (0,0.51) {\scriptsize ${#1}$};
        \node[color=black] at (0,0) {\scriptsize ${#2}$};
    }
  }
}

\tikzset{
  pics/cylnnrb/.style n args={2}{
    code = { %
        \filldraw[color=red, densely dashed,fill=black!30, thick] circle (0.35);
        \filldraw[color=blue, densely dashed, fill=white, thick]  circle (0.17);
        \node[color=black] at (0,0.51) {\scriptsize ${#1}$};
        \node[color=black] at (0,0) {\scriptsize ${#2}$};
    }
  }
}
\tikzset{
  pics/cylnnrr0/.style n args={2}{
    code = { %
        \filldraw[color=red,fill=black!30, thick] circle (0.35);
        \filldraw[color=red, fill=white, thick]  circle (0.17);
        \node[color=black] at (0,0.51) {\scriptsize ${#1}$};
        \node[color=black] at (0,0) {\scriptsize ${#2}$};
    }
  }
}
\tikzset{
  pics/cylnnrr/.style n args={2}{
    code = { %
        \filldraw[color=red, densely dashed,fill=black!30, thick] circle (0.35);
        \filldraw[color=red, densely dashed, fill=white, thick]  circle (0.17);
        \node[color=black] at (0,0.51) {\scriptsize ${#1}$};
        \node[color=black] at (0,0) {\scriptsize ${#2}$};
    }
  }
}
\tikzset{
  pics/cylnnbb0/.style n args={2}{
    code = { %
        \filldraw[color=blue,fill=black!30, thick] circle (0.35);
        \filldraw[color=blue, fill=white, thick]  circle (0.17);
        \node[color=black] at (0,0.51) {\scriptsize ${#1}$};
        \node[color=black] at (0,0) {\scriptsize ${#2}$};
    }
  }
}
\tikzset{
  pics/cylnnbb/.style n args={2}{
    code = { %
        \filldraw[color=blue, densely dashed,fill=black!30, thick] circle (0.35);
        \filldraw[color=blue, densely dashed, fill=white, thick]  circle (0.17);
        \node[color=black] at (0,0.51) {\scriptsize ${#1}$};
        \node[color=black] at (0,0) {\scriptsize ${#2}$};
    }
  }
}
\tikzset{
  pics/cylnn1/.style n args={2}{
    code = { %
        \filldraw[color=black,fill=black!30, thick] circle (0.35);
        \filldraw[color=black, fill=white, thick] (0.075,0) circle (0.15);
        \node[color=black] at (0,0.51) {\scriptsize ${#1}$};
        \node[color=black] at (0,0) {\scriptsize ${#2}$};
        \node[cross=3pt, very thick] at (-0.225,0) {};
    }
  }
}
\tikzset{
  pics/cylnn1rr/.style n args={2}{
    code = { %
        \filldraw[color=red,fill=black!30, thick] circle (0.35);
        \filldraw[color=red, fill=white, thick] (0.075,0) circle (0.15);
        \node[color=black] at (0,0.51) {\scriptsize ${#1}$};
        \node[color=black] at (0,0) {\scriptsize ${#2}$};
        \node[cross=3pt, very thick] at (-0.225,0) {};
    }
  }
}
\tikzset{
  pics/cylnn1bb/.style n args={2}{
    code = { %
        \filldraw[color=blue,fill=black!30, thick] circle (0.35);
        \filldraw[color=blue, fill=white, thick] (0.075,0) circle (0.15);
        \node[color=black] at (0,0.51) {\scriptsize ${#1}$};
        \node[color=black] at (0,0) {\scriptsize ${#2}$};
        \node[cross=3pt, very thick] at (-0.225,0) {};
    }
  }
}

\tikzset{
  pics/disc2holes/.style n args={2}{
    code = { %
        \filldraw[color=black,fill=black!30, thick] circle (0.35);
        \filldraw[color=black, fill=white, thick] (-0.15,0) circle (0.1);
        \filldraw[color=black, fill=white, thick] (0.15,0) circle (0.1);
        \node[color=black] at (-0.15,0.1) {\scriptsize ${#1}$};
        \node[color=black] at (0.15,0.1) {\scriptsize ${#2}$};
    }
  }
}

\tikzset{
  pics/torus1hole/.style n args={2}{
    code = { %
\draw [thick, fill=black!30] (-0.6,0) to [out=85,in=95] (0.4,0) to [out=-95,in=-85] (-0.6,0);
\filldraw [color=white, fill=white] (-0.15,0.01) to [out=40,in=130] (0.15,0.01) to [out=-160,in=-20] (-0.15,0.01);
\draw [thick] (-0.2,0.05) to [out=-40,in=-130] (0.2,0.05);
\draw [thick] (-0.15,0.01) to [out=40,in=130] (0.15,0.01);
\filldraw[color=black, fill=white, thick] (-0.375,0) circle (0.125);
\node[color=black] at (-0.15,0.1) {\scriptsize ${#1}$};
\node[color=black] at (0.15,0.1) {\scriptsize ${#2}$};
    }
  }
}
\tikzset{
  pics/torus1pt/.style n args={1}{
    code = { %
\draw [thick, fill=black!30] (-0.6,0) to [out=85,in=95] (0.4,0) to [out=-95,in=-85] (-0.6,0);
\filldraw [color=white, fill=white] (-0.15,0.01) to [out=40,in=130] (0.15,0.01) to [out=-160,in=-20] (-0.15,0.01);
\draw [thick] (-0.2,0.05) to [out=-40,in=-130] (0.2,0.05);
\draw [thick] (-0.15,0.01) to [out=40,in=130] (0.15,0.01);
\node[cross=3pt, very thick] at (-0.375,0) {};
\node[color=black] at (-0.375,0.2) {\scriptsize ${#1}$};
    }
  }
}

\usepackage{amsmath,amssymb,tikz-cd}
\usetikzlibrary{patterns}

\tikzset{
  vert/.style   ={circle, fill, inner sep=1.3pt},
  edge/.style   ={thick},
  newedge/.style={line width=2.2pt, gray!70!black},
  legi/.style   ={thick, blue!70!black},
  legn/.style   ={thick, red!80!black},
  lab/.style    ={font=\small},
}
\newcommand{\blob}[1]{%
  \draw[thick, preaction={fill=white}, pattern=north east lines] #1 circle (0.42);}

\tikzset{
  vert/.style={circle, fill, inner sep=1.3pt},
  edge/.style={thick},
  lX/.style  ={thick, blue!70!black},   
  l6/.style  ={thick, red!80!black},    
  lo/.style  ={thick},                  
}
\newcommand{\leg}[4]{\draw[#1] (#2) -- ++(#3:0.6)
  node[pos=1, shift={(#3:0.2)}, font=\footnotesize]{#4};}
\newcommand{\threeright}[1]{\leg{lo}{#1}{50}{} \leg{lo}{#1}{0}{} \leg{lo}{#1}{-50}{}}
\newcommand{\tworight}[1]{\leg{lo}{#1}{50}{} \leg{lo}{#1}{-50}{}}

\tikzset{
  tg/.style     ={x=0.5cm, y=0.5cm, baseline=(current bounding box.center)},
  vert/.style   ={circle, fill, inner sep=0.9pt},
  edge/.style   ={semithick},
  newedge/.style={line width=1.6pt, gray!70!black},
  leg/.style    ={semithick},
  legi/.style   ={semithick, blue!70!black},
  legn/.style   ={semithick, red!80!black},
  sm/.style     ={font=\scriptsize, inner sep=1pt},
  mapto/.style  ={-{Stealth[length=3pt]}, semithick},
  hookto/.style ={{Hooks[left,length=2pt]}-{Stealth[length=3pt]}, semithick},
}

\title{Higher-point amplitudes and recursion relations from the worldsheet in complex Liouville string}
\author{Aleksandr Artemev}
\affiliation{Institute for Theoretical Physics, University of Amsterdam,\\ Science Park 904, 1098 XH Amsterdam, The Netherlands}
\emailAdd{a.artemev@uva.nl}

\abstract{We consider multipoint perturbative string amplitudes in genus zero in complex Liouville string theory (CLS) in the worldsheet approach. Based on explicit worldsheet calculations and analytic bootstrap, we first compute  correlators involving ground ring operators, which additionally depend on a choice of integration cycle on the moduli space. Using this result and higher equations of motion in Liouville theory, we calculate higher-point tachyon correlators with one degenerate field and explain what contributions were missed in previous attempts to apply this method in the literature. This allows us to derive a recursion formula, which agrees with the general recursion for amplitudes known from ``matrix model'' approach to CLS. We also propose an expression for correlators with ground ring insertions in terms of tautological classes on the moduli space of curves and comment on generalization of these results to the case of rational minimal strings.}

\begin{document}

\maketitle

\makeatletter
\g@addto@macro\bfseries{\boldmath}
\makeatother

\section{Introduction}

There exists a class of interesting solvable string backgrounds that are referred to as ``minimal strings'' (in older literature, the term ``Liouville gravity'' was also used). They are defined by the worldsheet CFT that typically consists of three sectors: Liouville theory, some additional CFT called ``matter'' and the usual BRST-ghost system, such that the total central charge vanishes. Conditions on the matter CFT are not fully understood yet; however, different natural choices work, such as a free boson (``$c=1$ string'', \cite{Collier:2026pxi}),  another copy of Liouville theory \cite{Collier:2024kmo}, timelike Liouville \cite{Collier:2023cyw}, Runkel-Watts type theory \cite{Biancotto:2026ocf} or A-series CFT minimal model (``rational minimal string'', \cite{Seiberg:2004at}). Typical physical operators in such string theories are ``tachyons'' --- primary fields from the matter sector combined together with a Liouville primary so that their total conformal dimension is equal to $1$.

Interest in these theories lies in their conjectural duality with ``matrix models''. More precisely, conjectures of this type state that perturbative tachyon amplitudes in such theories are evaluated by Chekhov-Eynard-Orantin topological recursion (TR) \cite{Eynard:2007kz}. Thus in principle they can be evaluated to arbitrary order in perturbation theory. Most of the examples of such string backgrounds were found in recent years, starting with establishing similar results for two-dimensional Jackiw-Teitelboim gravity \cite{Saad:2019lba}, but the oldest one, the ``rational minimal string'', was first proposed back in the 80's. 

The simplicity of answers predicted by the dual approach and the general recursive structure of the amplitudes are quite mysterious from the worldsheet string theory point of view. Arguments in favor of the duality are very different in each available example, and worldsheet calculations from first principles that could check the duality can only be performed in the simplest cases (for the most recent numerical results see e.g. \cite{Balthazar:2017mxh, Christian:2026maj} for $c=1$ string and \cite{Rodriguez:2025rte} for rational minimal strings). General understanding of the origin of these dualities is still missing.

The rational minimal string is the first example when amplitudes involving integration over moduli spaces of complex dimension 1 (tree-level 4-point \cite{Belavin:2006ex} and genus one 1-point \cite{Artemev:2022sfi} amplitudes) were evaluated analytically, providing non-trivial confirmations of the duality. It was possible due to two (closely connected) special properties, both related to the presence of degenerate Virasoro primaries in the matter theory. The first is that there exist physical operators (nontrivial BRST-cohomology classes) of non-standard ghost numbers. Perhaps the most important are ``ground ring operators'' of ghost number zero, introduced in \cite{Witten:1991zd}. When introduced in the worldsheet correlator together with the tachyons, such operators give rise to closed non-top degree forms on moduli spaces of curves; one can then define a ``string amplitude'' involving ground ring operators provided an additional choice of cycle to integrate this form over. The second property is ``higher equations of motion'' (HEM)   --- special operator equations in Liouville CFT \cite{Zamolodchikov:2003yb}. For tachyon amplitudes they allow one to reduce the integral over moduli space to boundary terms, if at least one of the tachyons is built from a degenerate operator in the matter theory. In rational minimal strings all tachyons satisfy this, so one might hope that HEM can explain the recursion known from the dual side in full generality. However, attempts to generalize the calculation to higher-point amplitudes hit some roadblocks and to date have not been successful. Even in simpler examples one needs to be careful about several issues, related to non-analyticity of this theory in parameters, and add some terms by hand to make answers from this approach work \cite{Aleshkin:2016snp}.

Another interesting example of a minimal string theory is the ``complex Liouville string'' (CLS), introduced in \cite{Collier:2024kmo}. From the topological recursion point of view it requires some modifications of the usual setting (due to infinite number of ramification points of the associated spectral curve, see \cite{Weller:2026hud}). Nevertheless, this theory is worth studying because the string amplitudes have rich and interesting analytic structure. In particular, through known limiting properties of Liouville theory correlators, duality conjecture for CLS leads to other examples of ``matrix model'' duals for theories like $c=1$ and Runkel-Watts strings. CLS contains degenerate operators, similar to rational minimal string, but it also has analyticity in parameters, which simplifies its study. It is thus a natural playground to examine ground ring amplitudes and associated recursive structures on the worldsheet, which is the focus of this paper. The former question is interesting independently: it allows one to probe higher-point kinematics on the worldsheet in a simpler way than tachyon correlators, since one needs to perform fewer integrations.

The main results of this paper are listed below:
\begin{enumerate}
    \item We study in detail the CLS string amplitudes with one ground ring operator and four tachyons. It turns out that periods of the corresponding $2$-form can be reconstructed from their analytic properties and exact values at special points, that can be computed using worldsheet methods. Explicit answers for different periods are given in \eqref{fixedGRperiod}, \eqref{sumofs}, \eqref{fixtachyonperiod} and \eqref{gr2formotherdivisor}. 
    \item Together with ``higher equations of motion'', the ground ring results are used to study tachyon five-point amplitude with one degenerate operator. We show that this method allows one to reproduce the topological recursion answer \eqref{5pfdeg}. An important part of this calculation (section \ref{sec-codim-2}) identifies finite contributions to the amplitudes coming from ``corners'' of the moduli space. These are what fix the apparent discrepancy between worldsheet and TR. 
    \item Generalization of these arguments allows us to find a recursive formula \eqref{hem-tach-general} for any tree-level $n$-point tachyon amplitude with one degenerate field, which agrees with the predictions of dual ``matrix model'' approach. The ``corner'' calculation used for tachyon amplitudes can also be used to rederive expressions for ground ring periods and generalize them to $n \geq 5$ tachyons and arbitrary boundary divisors, see \eqref{gr-period-general}.
    \item We also propose a dictionary to compute cohomology classes associated to ground ring insertions in terms of tautological classes (section \ref{sec: gr-inttheory-dictionary}). Perhaps this might lead to some interpretation for these operators in the topological recursion setting.
\end{enumerate}
The remainder of the paper is organised as follows. In section \ref{sec: prelims} we provide necessary background on the worldsheet CFT in CLS and its physical operators, as well as construction of amplitudes in this theory. We also list some results expected from the matrix model approach to CLS. In section \ref{sec: 1GR-4T} we study an example of the amplitude containing a ground ring insertion. We find an exact expression for this amplitude, based on exact worldsheet calculations and analytic constraints, and provide some checks of the obtained answers. In section \ref{sec: higherptHEM} we explain the HEM-based calculation of tachyon 5-point amplitude with one degenerate field and comment on its generalization to higher-point case and for rational minimal strings. In section \ref{sec: inttheorygr} we motivate and propose the intersection-theoretic formula for ground ring insertions in terms of tautological classes on $\overline{\mathcal{M}}_{g,n}$. We conclude in section \ref{sec: discussion} with discussing possible directions for future work.

\section{CLS: worldsheet and matrix model} \label{sec: prelims}
\subsection{Definition of the worldsheet theory}
\paragraph{Liouville theory. }
We start with introducing our notations and some basic facts about Liouville CFT. Its central charge is parametrized in a standard way:
\begin{equation}
c = 1 + 6 Q^2,\,Q = b+ b^{-1}.
\end{equation}
In the path integral formulation, Liouville theory, in principle, is well-defined only for real $b$. From the point of view of conformal bootstrap, however, it makes sense to analytically continue structure constants and OPE data to define the theory for any value of $b \notin i\mathbb{R}$ (imaginary $b$ line is a natural boundary of analyticity for DOZZ structure constants). In the following we will take this perspective. 

Primary fields $V_p$ are labeled by parameter $p$, such that their conformal dimensions $\Delta_p = \frac{Q^2}{4}-p^2$. $V_p$ and $V_{-p}$ are identified by ``reflection symmetry'' --- correlators with such fields are equal up to normalization-dependent constant. In the normalization where this constant equals $1$, the three-point function (structure constant) reads
\begin{equation}
C_{b}(p_1, p_2,p_3) = \frac{\Gamma_b(2Q) \Gamma_b(\frac{Q}{2} \pm p_1 \pm p_2 \pm p_3)}{\sqrt{2} \Gamma^3_b(Q) \prod \limits_{j=1}^3 \Gamma_b(Q \pm 2p_j)},
\end{equation}
where the product over all choices of signs $\pm$ is implied. 

Higher-point correlators can be evaluated using the operator product expansion. It takes the form
\begin{equation}
V_{p_1} (x) V_{p_2}(0) =  \frac{i}{2}\int \limits_C dp\,\rho_b(p) C_b(p_1,p_2,p) |x|^{2(\Delta_p - \Delta_{p_1}-\Delta_{p_2})}[V_p(0)],
\end{equation}
where $[V_p]$ denotes contribution of all descendants and
$$
\rho_b(p) = 4 \sqrt{2} \sin 2\pi b p \sin 2\pi b^{-1}p .
$$
If all considered fields are in the spectrum of the theory, which means $bp_i \in \mathbb{R}$, integration contour $C$ is also the line $bp \in \mathbb{R}$\footnote{This is the prescription proposed in \cite{Collier:2024kwt} for complex central charge. If $b \in \mathbb{R}$, spectrum corresponds to imaginary $p$ instead.}. For other values of $p_i$ correlators are defined via analytic continuation, as a result of which some poles of structure constant can cross this contour. Then one needs to account for the corresponding residues (``discrete terms'') separately. In particular, it happens for ``degenerate'' values $p_i= \frac{mb + nb^{-1}}{2} \equiv p_{m,n} ;\,m,n \in \mathbb{N}$, where continuous contribution to the OPE vanishes. The corresponding fields will also be denoted $V_{m,n}$.

\paragraph{CLS amplitudes.} In CLS, worldsheet CFT is given by two copies of Liouville theory with (in general, complex) central charge, together with BRST ghost system $\mathfrak{b}, \mathfrak{c}, \mathfrak{\overline{b}}, \mathfrak{\overline{c}}$ of central charge $-26$. For the total conformal anomaly to vanish, two Liouville factors must have central charge parameters $b$ and $-ib$. In the following we will assume $b \in e^{i\pi/4}\mathbb{R}$, which makes two Liouville factors ``complex conjugate'' to each other. In particular, each has central charge $c \in 13 + i \mathbb{R}$. We will denote primary fields in the sector with $c=1+6(b+b^{-1})^2$ as $V$ and in the other one as $\Phi$. We will also refer to the second copy of Liouville theory as ``matter'', having in mind possible generalization of some of the arguments that follow to different choices of this CFT.

Physical operators are defined as cohomologies of BRST charge
\begin{equation}
Q_{\text{BRST}} = \mathcal{Q} + \overline{\mathcal{Q}},\,\mathcal{Q} = \frac{1}{2\pi i}\oint dz \left(\mathfrak{c} (T_L+T_M) + \normord{\mathfrak{c}\,\partial \mathfrak{c\,b}} \right).
\end{equation}
The simplest ones are built of primary fields in both sectors and are of ghost number $2$; they are referred to as ``tachyons''
\begin{equation}
T_p \equiv  \mathfrak{c}\mathfrak{\overline{c}} U_p =\mathfrak{c}\mathfrak{\overline{c}} V_p \Phi_{ip} .
\end{equation}
Relation between Liouville and matter momenta comes from BRST-closedness of $T$, which demands that dimensions of $V$ and $\Phi$ add up to $1$. 

Tachyon amplitudes are built from correlators of $T_p$ with additional $\mathfrak{b}$-ghost insertions, integrated over the moduli space $\mathcal{M}_{g,n}$. In particular, in genus zero $n$-point amplitudes, which we will focus on in this work, we can choose $n-3$ cross-ratios as coordinates on $\mathcal{M}_{0,n}$ and the role of $\mathfrak{b}$-ghost insertions reduces to erasing $\mathfrak{c}$-ghost dressing for $(n-3)$ operators, turning them into their integrated form $U_p$. Genus zero amplitudes then read
\begin{equation}
\mathsf{A}_{0,n} = C_{S_2}^{(b)} \prod \limits_{j=1}^n \mathcal{N}(p_j)  \int\left(\prod \limits_{k=1}^{n-3} d^2z_k\right) \left\langle \prod \limits_{k=1}^{n-3} U_{p_k}(z_k) \times T_{p_{n-2}}(0) T_{p_{n-1}}(1) T_{p_{n}}(\infty) \right\rangle .
\end{equation}
Here we introduced the ``sphere partition function'' and normalization factors
\begin{equation}
C_{S_2}^{(b)} = 32 \pi^4 \left(\frac{\sin \pi b^2 \sin \pi b^{-2}}{b^2-b^{-2}} \right)^2,\,\mathcal{N}(p)= - \frac{(b^2 - b^{-2})\rho_b(p)}{8\pi \sqrt{2}\,p\,\sin \pi b^2 \sin \pi b^{-2}} .
\end{equation}
 $\mathcal{N}(p)$ are needed only for convenience and to make comparison to dual approach more straightforward. For example, in this normalization the three-point amplitude $\mathsf{A}_{0,3}$ becomes a doubly-periodic function of all momenta with periods $b, b^{-1}$. ``Canonically normalized'' tachyon operators will be denoted as $\mathcal{T}_p = \mathcal{N}(p) T_p,\,\mathcal{U}_p = \mathcal{N}(p) U_p$.
\subsection{Ground ring operators}
In CLS, similar to minimal string, matter theory has degenerate primary operators $\Phi_{m,n}$ with a null descendant $D_{m,n}\Phi_{m,n}$ at level $mn$. This leads to the existence of a discrete family of BRST-closed operators of ghost number zero, called the ``ground ring'' (GR) \cite{Witten:1991zd}. They have the general form
$$
O_{m,n} = H_{m,n} \overline{H}_{m,n} V_{m,n} \Phi_{m,n} ,
$$
where $H_{m,n} = L_{-1}^{mn-1} + \dots$ is a polynomial of degree $mn-1$ built from Virasoro modes $L_{-k}, M_{-k}$ (in Liouville and matter sector, respectively) and ghost modes $\mathfrak{b}_{k},\mathfrak{c}_{k}$. They are constructed from descendants in such a way that action of BRST charge yields precisely the combination of null descendants in Liouville and matter theories; they are thus nontrivial representatives of BRST cohomology. No general formula for $H_{m,n}$ is known, but they can be found from the requirement of BRST-closedness \cite{imbimbo1992}; the first few examples are
\begin{align*}
    &H_{2,1} =  L_{-1} - M_{-1} + b^2 \mathfrak{b}_{-2} \mathfrak{c}_1,\\
    &H_{3,1} =L^2_{-1} - L_{-1}M_{-1} + M^2_{-1} + 2b^2(2b^2 + 1)L_{-2} + 2b^2(2b^2- 1)M_{-2}+ \\
    &+(2b^2(L_{-1}-M_{-1})-4b^4(L_{-1}+M_{-1}))\mathfrak{b}_{-2} \mathfrak{c}_1
\end{align*}
and similar expressions for $H_{1,2}, H_{1,3}$ with $b$ replaced by $b^{-1}$. We also introduce operator $R_{m,n}$:
\begin{equation}
R_{m,n} \equiv \mathfrak{b} H_{m,n}\text{ s.t. }\partial O_{m,n} = Q_{\text{BRST}} (\mathfrak{b}O_{m,n}) =  Q_{\text{BRST}} (R_{m,n} \overline{H}_{m,n} V_{m,n}\Phi_{m,n})
\end{equation}
(the equation is valid up to null descendants). This equation reflects independence of the correlator on ground ring position.

Similar to tachyons, it will be more convenient to consider a ``canonically normalized'' ground ring operator
\begin{equation}
    \mathcal{O}_{m,n} \equiv - \frac{\pi \mathcal{N}(\frac{mb}{2}-\frac{n}{2b})}{B_{m,n}} O_{m,n},
\end{equation}
where $B_{m,n}$ will be introduced later. In this normalization ground ring operator has special properties \cite{Belavin:2006ex}: simple OPE with (generic) tachyons
\begin{equation}
\mathcal{O}_{m,n}(x) \mathcal{T}_p(0) = \sum \limits_{k=1-m:2}^{m-1} \sum \limits_{l=1-n:2}^{n-1} \mathcal{T}_{p + \frac{kb}{2} + \frac{l b^{-1}}{2}}(0) + \dots
\end{equation}
where $:2$ denotes a sum with step $2$, and among themselves
\begin{equation}
 \mathcal{O}_{m,n}(x) \mathcal{O}_{r,s}(0) = \sum \limits_{k=|m-r|+1:2}^{m+r-1} \sum \limits_{l=|n-s|+1:2}^{n+s-1} \mathcal{O}_{k,l}(0) + \dots
\end{equation}
The sum is over allowed fusion channels in $[\Phi_{m,n}][\Phi_{r,s}]$ OPE. To rephrase, ground ring operators obey the same algebra as Chebyshev polynomials of the second kind: $\mathcal{O}_{r,s} = \mathbf{U}_{r-1} (\mathcal{O}_{2,1}/2) \mathbf{U}_{s-1} (\mathcal{O}_{1,2}/2)$.
\paragraph{Ground ring correlators. } General string-theoretic prescription \cite{Polchinski:1998rq} produces closed differential forms on moduli space of curves $\mathcal{M}_{g,n}$ (or, more precisely, on the space $P_{g,n} \to \mathcal{M}_{g,n}$, which parametrizes complex curves together with the choice of local coordinates near the punctures; see, e.g., \cite{Sen:2024nfd}) from correlation functions of any BRST-invariant operators. However, the degree of this form depends on the total ghost number. If every operator is a tachyon of standard ghost number $2$, then we get the top degree form which we can integrate over the whole moduli space and get a number --- the string amplitude. Introducing ghost number zero operators, such as $\mathcal{O}_{m,n}$, reduces the degree by $2$ for each such insertion. For example, if we consider $(n-3)$ ground ring operators and $3$ tachyons, the result is a closed 0-form, or constant function on $\mathcal{M}_{0,n}$, which does not require any integrations. The main example in this paper is the case with one ground ring and 4 tachyons (1GR-4T), which, according to this discussion, produces a ``ground ring 2-form'' on $\mathcal{M}_{0,5}$. In standard cross-ratio coordinates it reads
\begin{align}
 \Omega_{1,s}&=\langle  \mathcal{T}_{p_{1}}(0) O_{1,s}(x) \mathcal{U}_{p_2}(y) \mathcal{T}_{p_{4}}(1) \mathcal{T}_{p_{5}}(\infty)  \rangle dy\wedge d\overline{y} +\nonumber \\
 &+ \langle  \mathcal{T}_{p_{1}}(0) R_{1,s}\overline{H}_{1,s} V_{1,s} \Phi_{1,s}(x) \left(\mathfrak{c}\mathcal{U}_{p_2}\right)(y) \mathcal{T}_{p_{4}}(1) \mathcal{T}_{p_{5}}(\infty)  \rangle dx\wedge d\overline{y} +\nonumber \\
 &+ \langle  \mathcal{T}_{p_{1}}(0) \overline{R}_{1,s}H_{1,s} V_{1,s} \Phi_{1,s}(x) \left(\overline{\mathfrak{c}} \mathcal{U}_{p_2}\right)(y) \mathcal{T}_{p_{4}}(1) \mathcal{T}_{p_{5}}(\infty)  \rangle dy \wedge d\overline{x} -\nonumber \\
 &- \langle  \mathcal{T}_{p_{1}}(0) R_{1,s}\overline{R}_{1,s} V_{1,s} \Phi_{1,s}(x) \mathcal{T}_{p_2}(y) \mathcal{T}_{p_{4}}(1) \mathcal{T}_{p_{5}}(\infty)  \rangle dx\wedge d\overline{x}. \label{gr2form}
\end{align}
To obtain a number we need to choose a 2-cycle over which to integrate this form. Thus, several a priori different numbers --- periods of this form --- can be called the ``ground ring correlator'', depending on this additional choice. 
The fact that this form is closed can be understood by using the identities for BRST-descendants
\begin{align*}
    \partial H_{1,s} V_{1,s}\Phi_{1,s} &= \mathcal{Q} \left(R_{1,s} V_{1,s} \Phi_{1,s} \right), \\
    \partial (\mathfrak{c} \mathcal{U}_p) &= \mathcal{Q} \left(\mathcal{U}_p \right),
\end{align*}
and moving the action of BRST charge from one operator to the other inside the correlator.
\paragraph{Symmetry in parameters. } For tachyon amplitudes on the sphere result does not depend on which $n-3$ operators are taken in the integrated form $\mathcal{U}$; in particular, it is symmetric in permutations of all tachyon parameters. This can be shown by using transformation properties of CFT correlators under global conformal group. For some fixed period of ground ring 2-form the same symmetry is not expected: the integrated tachyon plays a distinguished role (although we will see that in CLS some periods, unexpectedly, do turn out to be symmetric in tachyon permutations for 1 ground ring and 4 tachyons case).

However, one can check straightforwardly that ``ground ring 2-form'' \eqref{gr2form} is always symmetric in the following sense. Assume, for example, that we want to compute a correlator with a different integrated tachyon $\mathcal{U}_{p_4}$ instead of $\mathcal{U}_{p_2}$. For that we need to integrate the $2$-form
\begin{equation}
\langle  \mathcal{T}_{p_{1}}(0) O_{m,n}(w) \mathcal{U}_{p_4} (t) \mathcal{T}_{p_{2}}(1) \mathcal{T}_{p_{5}}(\infty)  \rangle dt \wedge d\overline{t} \label{otherinttach}
\end{equation}
over a cycle in $\mathcal{M}_{0,5}$ with fixed $w$. If we rewrite \eqref{gr2form} in coordinates $w = \frac{x}{y}, t = \frac{1}{y}$, using conformal properties of the correlator, the $dt\wedge d\overline{t}$-term will be given precisely by \eqref{otherinttach}. For examples such as $O_{1,2}$ this can be confirmed by an explicit calculation: all terms of \eqref{gr2form} ($dy\wedge d\overline{x}, dx \wedge d\overline{y}, \dots$) need to be accounted for to reproduce \eqref{otherinttach} in $(w,t)$ coordinates.

\subsection{Higher equations of motion}
The presence of degenerate fields and ground ring operators allows us to make use of a different interesting structure in Liouville field theory --- the so-called ``higher equations of motion'' (HEM).  In the modern approach to Liouville CFT they were first discussed in detail by Al. Zamolodchikov \cite{Zamolodchikov:2003yb}. Introduce first the ``logarithmic operators'':
\begin{equation}
V'_p = \frac{1}{2} \frac{\partial}{\partial p} V_p.
\end{equation}
These are not conformal primary fields: $L_0$ acts on them non-diagonally. We will also write $V'_{m,n}$ for $V'_p$ evaluated at $p=p_{m,n}$.

The higher equations of motion state that in any Liouville correlator one can replace
\begin{equation}
 V_{m,-n} = B_{m,n}^{-1} D_{m,n} \overline{D}_{m,n} V'_{m,n} ,\label{liouv-hem}
\end{equation}
where $B_{m,n}$ can be computed in terms of DOZZ structure constant as
\begin{equation}
B_{m,n} = \frac{\prod\limits_{k=1-m:2}^{m-1} \prod\limits_{l=1-n:2}^{n-1} (p_1 \pm p_2 - p_{k,l})^2 }{C_b(p_{m,-n},p_1,p_2)} \frac{1}{2} \frac{\partial}{\partial p} \Bigg|_{p=p_{m,n}} C_b(p,p_1,p_2)
\end{equation}
and the result is independent of $p_1,p_2$. Note that $V_{m,-n}$ is precisely the ``dressing'' for degenerate field $\Phi_{m,n}$ in the matter sector: they combine into tachyon operator $U_{m,-n}$ in the full string theory. In this context, HEM can be promoted to a ``cohomological'' version:
\begin{equation}
    U_{m,-n}  = B_{m,n}^{-1} (\partial H_{m,n} - \mathcal{Q} R_{m,n}) (\overline{\partial H}_{m,n} - \overline{\mathcal{Q} R}_{m,n}) V'_{m,n} \Phi_{m,n} .
\end{equation}
This can be understood by noting that if there was no derivative over momentum, operator $(\partial H_{m,n} - \mathcal{Q} R_{m,n})$ would produce null descendants in both Liouville and matter sectors. For logarithmic operator the result is $D_{m,n} V'_{m,n} \Phi_{m,n}$ up to a correction, which is nullified by the action of anti-holomorphic operator $(\overline{\partial H}_{m,n} - \overline{\mathcal{Q} R}_{m,n})$.

In canonical normalization this equation takes a particularly simple form: the integrated tachyon is rewritten as a total derivative
\begin{equation}
    \mathcal{U}_{m,-n}  = - \frac{1}{\pi} \partial \overline{\partial} \mathcal{O}'_{m,n} + \text{BRST-exact},\,\label{com-hem-4pt}
\end{equation}
where $\mathcal{O}'$ is the logarithmic partner of $\mathcal{O}$, i.e., it has an additional derivative over momentum. There is a different way to write this equation, which is to represent a tachyon operator $\mathcal{T}_{m,-n}$ as a BRST-descendant $\mathcal{Q}\overline{\mathcal{Q}}\mathcal{O}'_{m,n}$. This agrees with the general lore that introducing BRST-exact insertions in string theory produces total derivative terms on the moduli space. Usually it means that they completely decouple; however, here we needed to introduce logarithmic operators which are not regular conformal fields and the standard argument fails. Near the boundary of moduli space correlators with $\mathcal{O}'_{m,n}$ have logarithmic behavior, which leads to nontrivial contributions to the integral.

\paragraph{Review of HEM for four-point amplitude. }This observation was used to compute tachyon four-point function on the sphere with degenerate tachyons in rational minimal string \cite{Belavin:2006ex} and further in CLS \cite{Collier:2024kwt}. First of all, $\mathcal{Q}$-exact terms in \eqref{com-hem-4pt} do not matter in this case and we can use HEM in this form. What is needed to compute the boundary terms, where $\mathcal{O'}$ collides with the other tachyon operators, are logarithmic terms in $\mathcal{O'T}$-OPE. They can be evaluated and read
\begin{multline}
\mathcal{O}'_{m,n}(x) \mathcal{T}_p(0) = \log |x|^2 \sum \limits_{k=1-m:2}^{m-1} \sum \limits_{l=1-n:2}^{n-1} q^{m,n}_{k,l}(p)\mathcal{T}_{p + \frac{kb}{2} + \frac{l b^{-1}}{2}}(0) + \dots,\,\\q^{m,n}_{k,l}(p) = \frac{mb+nb^{-1}}{2} - |p+\frac{kb}{2}+\frac{lb^{-1}}{2}| .\label{Oprime-T-OPE}
\end{multline}
The $q$-factor, together with the coordinate logarithm, arises from differentiating the power factor in the OPE with respect to Liouville momenta.  As explained in \cite{Collier:2024kwt}, by $|p+\frac{kb}{2}+\frac{lb^{-1}}{2}|$ in CLS one actually should mean some branch of $\sqrt{(p+\frac{kb}{2}+\frac{lb^{-1}}{2})^2}$ function. We will use the simpler $|\dots |$ notation, as in \cite{Belavin:2006ex}, from now on. Since the result of the OPE is again a tachyon operator, corresponding boundary term will be proportional to a three-point tachyon amplitude with shifted momentum.

There is one more contribution to the four-point amplitude, which in the standard derivation comes from the ``vicinity of infinity''. Cross-ratio $x$ becomes a bad coordinate at this point, so we need to pass to a different chart $w=\frac{1}{x}$. Logarithmic operator $\mathcal{O}'_{m,n}$ transforms in a non-homogeneous way under such coordinate transformations, with an additive term proportional to $\mathcal{O}_{m,n}$. Overall to compute this contribution we can replace
\begin{equation}
\mathcal{O}'_{m,n}(x) \to 2p_{m,n} \log |x|^2 \mathcal{O}_{m,n}(\infty),\,x \to \infty .
\end{equation}
This term is then proportional to the correlator $\langle \mathcal{O}_{m,n} \mathcal{T}\mathcal{T}\mathcal{T}\rangle$, which can be expressed as a sum of shifted tachyon three-point amplitudes, using $\mathcal{OT}$-OPE. In fact, this contribution and $\frac{mb+nb^{-1}}{2}$ terms in \eqref{Oprime-T-OPE} should be seen as a single ``curvature term'', proportional to degenerate Liouville momentum $p_{m,n}$. As we will see later, in higher tachyon amplitudes such contributions will also appear after applying HEM and it will be easier to interpret them all together. 

If we additionally use shift invariance properties of $\mathsf{A}_{0,3}$, the answer for four-point amplitude in CLS obtained using this method takes the form 
\begin{multline}
\mathsf{A}_{0,4}(p_1,p_2,p_3,\frac{mb-nb^{-1}}{2}) = \mathsf{A}_{0,3}(p_1+p_{m,n}-\frac{Q}{2},p_2,p_3) \times \\\times \left(mn\, p_{m,n} - \sum \limits_{i=1}^3 \sum \limits_{k=1-m:2}^{m-1} \sum \limits_{l=1-n:2}^{n-1} |p_i + p_{k,l}| \right).
\end{multline}
Note that, of course, in CLS Liouville and matter theory play a symmetric role, so a similar line of reasoning allows us to compute tachyon amplitude with degenerate field in Liouville sector $\mathsf{A}_{0,4}(p_1,p_2,p_3,\frac{mb+nb^{-1}}{2})$.

\paragraph{Higher-point amplitudes. } Cohomological form of higher equations of motion suggests that all tachyon correlators with one degenerate field should be reducible to boundary terms and, in principle, computable. In particular, in rational minimal string degenerate tachyons is all we have and  one could hope that this approach (with proper modifications)  computes any correlator. Previous attempts to extend it to amplitudes with more than one-dimensional moduli spaces, however, missed some contributions, which led to disagreement with the predictions of the dual ``matrix model''. This is the gap we intend to fix in this work by considering the situation more carefully.

Let us describe more precisely what happens when we apply HEM for the example of five-point tachyon correlator. We have
\begin{equation}
\int d^2x\,d^2y\left\langle \mathcal{U}_{m,-n}(x) \mathcal{U}_{p_1}(y)\mathcal{T}_{p_2}(0) \mathcal{T}_{p_3}(1) \mathcal{T}_{p_4}(\infty) \right\rangle
\end{equation}
and we substitute $U_{m,-n} = B_{m,n}^{-1} (\partial H - \mathcal{Q}R) \overline{(\dots)} V'_{m,n} \Phi_{m,n}$. At this stage the answer is manifestly symmetric in permutations of $p_1, \dots, p_4$. 

For the four-point function we could have ignored the $\mathcal{Q}$-exact terms. Here they do contribute because the leftover correlator is only $\mathcal{Q}$-invariant up to a total derivative: $\mathcal{Q}U_p = \partial (\mathfrak{c} U_p)$. In particular, it was noted in \cite{Artemev:2022rng}  that these terms ``fix'' the OPE of $O'$ with $U$, so that it takes the same form as OPE with $T$. 

Moving the BRST charge to the integrated field we see that the integrand is given by a sum of 4 pieces
\begin{align}
\frac{\mathcal{N}(\frac{mb}{2}-\frac{n}{2b})}{B_{m,n}} \Bigl(&\frac{\partial}{\partial x} \frac{\partial}{\partial \overline{x}}\langle  O'_{m,n}(x) \mathcal{U}_{p_1}(y) \mathcal{T}_{p_{2}}(0) \mathcal{T}_{p_{3}}(1) \mathcal{T}_{p_{4}}(\infty)  \rangle  \nonumber \\
 &+ \frac{\partial}{\partial y} \frac{\partial}{\partial \overline{x}}\langle  R_{m,n} \overline{H}_{m,n} \Theta'_{m,n}(x) \mathfrak{c}\mathcal{U}_{p_1}(y) \mathcal{T}_{p_{2}}(0) \mathcal{T}_{p_{3}}(1) \mathcal{T}_{p_{4}}(\infty)  \rangle  \nonumber \\
 &+ \frac{\partial}{\partial x} \frac{\partial}{\partial \overline{y}}\langle \overline{R}_{m,n} H_{m,n} \Theta'_{m,n}(x) \overline{\mathfrak{c}} \mathcal{U}_{p_1}(y) \mathcal{T}_{p_{2}}(0) \mathcal{T}_{p_{3}}(1) \mathcal{T}_{p_{4}}(\infty)  \rangle \nonumber\\
 &-  \frac{\partial}{\partial y} \frac{\partial}{\partial \overline{y}} \langle  R_{m,n} \overline{R}_{m,n} \Theta'_{m,n}(x)  \mathcal{T}_{p_1}(y) \mathcal{T}_{p_{2}}(0) \mathcal{T}_{p_{3}}(1) \mathcal{T}_{p_{4}}(\infty)   \rangle  \Bigr),\label{loggr2form}
\end{align}
where $\Theta'_{m,n} = V'_{m,n} \Phi_{m,n}$. In other words, we have a total derivative of the logarithmic analogue of ground ring 2-form (cf. \eqref{gr2form}).
\begin{equation}
\left\langle \mathcal{U}_{m,-n}(x) \mathcal{U}_{p_1}(y) \mathcal{T}_{p_2}(0) \mathcal{T}_{p_3}(1) \mathcal{T}_{p_4}(\infty) \right\rangle = -\frac{1}{\pi} \partial \overline{\partial} \Omega^{\text{log}}_{m,n} . \label{HEM2form}
\end{equation}
This formulation suggests that HEM calculation will, as before, reduce the problem to computing correlators with the ground ring. We will see, however, that unlike the four-point case knowing the correlator with the unintegrated ground ring operator will not be enough for this computation. In the next section we will study how one can compute different periods of the GR $2$-form in CLS. 
\subsection{Expected answers from dual approach in CLS}
Here for future reference we collect some predictions of topological recursion approach to CLS for tachyon amplitudes. For simplicity here and below we will always assume that parameters of degenerate fields $r$ and $s$ are odd; even case would require half-period shift in some expressions, which makes them less convenient to analyze. First, we have a general intersection-theoretic formula of \cite{Collier:2024lys}, with amplitude represented as a sum over stable graphs of genus $g$ with $n$ legs:
\begin{multline}
\mathsf{A}_{g,n} = \int \limits_{\overline{\mathcal{M}}_{g,n}}\sum \limits_{\Gamma} \frac{1}{|\text{Aut\,}\Gamma|} \iota_{\Gamma *} \Theta_\Gamma,\, \\ \Theta_\Gamma = \prod \limits_e \left(\int' -2q_e dq_e\,e^{-q_e^2 (\psi_{e\circ} + \psi_{e\bullet})} \right) \prod \limits_{l} e^{-p_l^2 \psi_l} \prod \limits_v \mathcal{N}_v\, e^{\sum \limits_{i=1}^\infty t_i \kappa_i^v}.
\end{multline}
Cohomology class $\Theta_\Gamma$ is given in terms of tautological classes (see \cite{Zvonkine2012AnIT} for an introduction) as a product over edges $e$, legs $l$ and vertices $v$ of the graph. With each edge a momentum $q_e$ is associated, which we integrate over; $\psi_{e\circ}$ and $\psi_{e\bullet}$ are two psi-classes associated to half-edges of $e$. Integrand is understood as a power series in $\psi$-classes, where for each graph only finitely many terms contribute, and at each order the integral is evaluated according to the rule
\begin{equation}
\int' dq\,q^{2k+1} e^{-aq} \equiv \frac{\Gamma(2k+2)}{a^{2k+2}} \delta_{a\neq 0}.
\end{equation}
For each vertex $v$ we have TQFT factors
\begin{equation}
\mathcal{N}_v = b^{2g_v-2+n_v} \sum \limits_{m=1}^\infty (-1)^{m(2g_v-2+n_v)}\frac{\prod \limits_{p_\alpha \in h_v} \sqrt{2} \sin 2\pi m b p_\alpha}{(\sqrt{2} \sin \pi m b^2)^{2g_v-2+n_v}},
\end{equation}
where the product is over all momenta $p_\alpha \in h_v$ on edges and legs that enter $v$, and an exponential of $\kappa$-classes with coefficients
\begin{equation}
\sum \limits_{i=1}^\infty t_i \kappa_i = \frac{b^2 + b^{-2}}{4} \kappa_1 - \sum \limits_{k=1}^\infty \frac{B_{2k}}{2k(2k)!} \kappa_{2k}.
\end{equation}
At each vertex of the stable graph  $\kappa$- and $\psi$-classes combine in the same inhomogeneous class that computes amplitudes in Virasoro minimal string, or ``quantum volumes'' $\mathsf{V}_{g,n}$ \cite{Collier:2023cyw}. For this reason these volumes enter explicit formulas below.

\paragraph{Three- and four-point amplitudes. } These were written down explicitly in \cite{Collier:2024kmo}. Introduce
\begin{equation}
\mathcal{N}_{S}(m) = \prod \limits_{s \in S} \sin 2\pi m b p_s,\,G_k(m,n) = \left(\frac{\delta_{m \neq n}}{(m-n)^{2k}} -  \frac{1}{(m+n)^{2k}} \right).
\end{equation}
Then the three-point amplitude in genus zero is
\begin{equation}
\mathsf{A}_{0,3} = 2b \sum \limits_{m=1}^\infty (-1)^m\frac{\mathcal{N}_{123}(m)}{\sin \pi m b^2}
\end{equation}
and four-point amplitude reads
\begin{align}
    \mathsf{A}_{0,4}^{(b)}(\boldsymbol{p})=2b^2 \mathsf{V}_{0,4}(i\boldsymbol{p}) \mathsf{C}(\boldsymbol{p})+\mathsf{E}(\boldsymbol{p})\ ,
\end{align}
where
\begin{align}
    \mathsf{C}(\boldsymbol{p})&=\sum_{m=1}^\infty \frac{\mathcal{N}_{1234}(m)}{\sin(\pi m b^2)^2}\ , \\
    \mathsf{E}(\boldsymbol{p})&\equiv\sum \limits_{ab|cd} \mathsf{E}_{ab|cd} =\frac{1}{\pi^2}\sum \limits_{ab|cd}\sum_{m_1,m_2=1}^\infty (-1)^{m_1+m_2}\frac{\mathcal{N}_{12}(m_1)\mathcal{N}_{34}(m_2)}{\sin(\pi m_1 b^2)\sin(\pi m_2 b^2)} G_1(m_1,m_2)\ 
\end{align}
\paragraph{Five-point amplitude}
The five-point amplitude in genus zero can be decomposed according to the contributing stable graphs (i.e. by the number of internal edges)
\begin{equation}
\mathsf{A}_{0,5}^{(b)}= 2b^3 \mathsf{V}_{0,5} C_5 + \sum \limits_{ab|cde}  \left(\mathsf{V}_{0,4}(ip_{cde},0)\text{1E}^{(1)}_{cde} + \text{1E}^{(2)}_{cde} \right) + \sum \limits_{ab|c|de}\text{2E}_{ab|c}.
\end{equation}
$a,b,c,d,e$ are different orderings of the set $\lbrace1,\dots,5\rbrace$. The factors $C_5, \text{1E}^{(k)},\text{2E}$ are trigonometric ($b^{-1}$-periodic):
\begin{align}
& C_5 = \sum \limits_{m \geq 1} (-1)^m \frac{\mathcal{N}_{12345}(m)}{\sin^3 \pi m b^2} , \nonumber \\
&  \text{1E}^{(1)}_{cde} = \frac{b}{\pi^2} \sum \limits_{m,n} (-1)^m \frac{\mathcal{N}_{ab}(m) \mathcal{N}_{cde}(n)}{\sin \pi m b^2 \sin^2 \pi n b^2} G_1(m,n),  \nonumber \\
& \text{1E}^{(2)}_{cde} = \frac{3}{2\pi^4 b} \sum \limits_{m,n} (-1)^m \frac{\mathcal{N}_{ab}(m) \mathcal{N}_{cde}(n)}{\sin \pi m b^2 \sin^2 \pi n b^2} G_2(m,n), \nonumber \\
& \text{2E}_{ab|c} = \frac{1}{2\pi^4 b} \sum \limits_{m,n,k} (-1)^{m+n+k} \frac{\mathcal{N}_{ab}(m) \mathcal{N}_{c}(n)\mathcal{N}_{de}(k)}{\sin \pi m b^2 \sin \pi n b^2 \sin \pi k b^2} G_1(m,n) G_1(n,k).\label{5pf-ingredients}\end{align}
When specialized to degenerate momenta, this formula becomes
\begin{align}
    \mathsf{A}_{0,5}^{(b)}(\boldsymbol{p},\tfrac{b}{2}+\tfrac{s}{2b})
    &= \frac{s-b^2}{b} \left(s\mathsf{A}_{0,4}^{(b)}(\boldsymbol{p}) - \frac{s(s^2-1)}{4} \mathsf{C}(\boldsymbol{p}) \right) + \frac{(s^2-1)^2}{16b} \mathsf{C}(\boldsymbol{p}) - \frac{s^2-1}{2b} \mathsf{E}(\boldsymbol{p})-\nonumber\\
    &- \sum_{i=1}^4\sum_{\ell = (1-s)/2}^{(s-1)/2} |p_i-b^{-1} \ell| \mathsf{A}_{0,4}^{(b)}(\boldsymbol{p} \setminus p_i,p_i-b^{-1}\ell)\ . \label{5pfdeg}
\end{align}
\section{One ground ring-four tachyon correlator} \label{sec: 1GR-4T}
In this section the goal is to find an expression for
\be 
\mathsf{A}_{0,4|1}(p_4,p_1,p_2,p_3|(r,s))=\int \d^2 z \langle \mathcal{O}_{r,s}(x) \mathcal{T}_{p_1}(z_1)\mathcal{T}_{p_2}(z_2)\mathcal{T}_{p_3}(z_3)
\mathcal{U}_{p_4}(z) \rangle\ ,\label{gr2formbasicperiod}
\ee
i.e., the fixed ground ring period of GR 2-form $\Omega$. This object should be independent of $x$, as well as $z_i$. For $(r,s)=(1,1)$, the ground ring operator is the identity and the correlator reduces to the usual tachyon four-point function. 

\subsection{Expected analytic properties and the ansatz}
We start with discussing discontinuities of this correlator as a function of tachyon momenta. 

The analysis is exactly parallel to the case of tachyon correlators in \cite{Collier:2024kwt}: discontinuities are associated with discrete terms appearing in Liouville OPE. Unlike the continuum part, these discrete terms can give divergent contributions to the integral when $z \to z_i$; this requires analytic continuation that is not single-valued.

This brings cuts in the complex plane of $p_i \pm p_j$ starting at 
\begin{equation}
\sigma_i p_i + \sigma_j p_j + (k+\frac{1}{2}) b + (l+\frac{1}{2})b^{-1} \equiv p_* = 0,\quad k,l \in \mathbb{Z}.
\end{equation}
for all signs $\sigma_i,\sigma_j = \pm 1$. Discontinuities are
\be 
\Disc_{p_*=0}\mathsf{A}_{0,4|1}(p_4,p_1,p_2,p_3|(r,s))=8\pi i \Res_{p=p_*} p\mathsf{A}_{0,3|1}(p_1,p_2,p|(r,s)) \mathsf{A}_{0,3}(p_3,p_4,p)+\text{2 perms.}
\ee
Moreover,
\be 
\mathsf{A}_{0,3|1}(p_1,p_2,p_3|(r,s))=rs \mathsf{A}_{0,3}(p_1,p_2,p_3+\frac{(r-1)b}{2}+\frac{(s-1)}{2b})=rs \mathsf{A}_{0,3}(p_1,p_2,p_3)\ ,
\ee
since this can be obtained simply by applying the OPE of the ground ring operator with one of the tachyons. We used that $\mathsf{A}_{0,3}$ is doubly periodic and in the last step that $r$ and $s$ are odd. Note that discontinuities are only associated with degenerations of the surface when $p_4$ ``bubbles off'' with one of the $p_i$: since we only integrate over $z$, the position of $\mathcal{U}_{p_4}$, degeneration where $p_4$ and two of the stationary points bubble off cannot happen. Hence cuts and discontinuities are the same as for the four-point amplitude $\mathsf{A}^{(b)}_{0,4}(p_4,\vec{p})$ up to the factor $rs$.

Poles in $p_i$ come from the poles of moduli space integrand, which appear if screening condition in one of the Liouville factors is satisfied, i.e. when
\be 
\pm \Big(\frac{rb}{2}+\frac{s}{2b}\Big) \pm p_1 \pm p_2 \pm p_3 \pm p_4 \in \Big(\frac{3}{2}+\NN_0\Big) b+ \Big(\frac{3}{2}+\NN_0\Big) b^{-1}
\ee
or
\be 
\pm \Big(\frac{rb}{2}-\frac{s}{2b}\Big) \pm p_1 \pm p_2 \pm p_3 \pm p_4 \in \Big(\frac{3}{2}+\NN_0\Big) b- \Big(\frac{3}{2}+\NN_0\Big) b^{-1}\ .
\ee
When $r$ and $s$ are odd, this means that the signed sum of momenta $p_1,\dots,p_4$ lies in the lattice $\ZZ b+\ZZ b^{-1}$. In total, the poles are the same as for the four-point amplitude $\mathsf{A}^{(b)}_{0,4}(p_4,\vec{p})$ (or at the very least there are no others). 

All of this brings us to the following ansatz for 1GR-4T amplitude \eqref{gr2formbasicperiod}:
\begin{equation}
\mathsf{A}_{0,4|1}(p_4,\vec{p}|(r,s)) = rs\mathsf{A}^{(b)}_{0,4}(p_4,\vec{p}) + \delta \mathsf{A}(p_4,\vec{p}|r,s).
\end{equation}
The first term accounts for necessary discontinuities; thus the correction $\delta \mathsf{A}$ must be a meromorphic function of momenta.

There is a subtlety in the above discussion that is necessary to address. For four-point function $\mathsf{A}^{(b)}_{0,4}(p_4,\vec{p})$ the cuts start at, e.g.,
\begin{align*}
&\pm p_1 \pm p_4 = mb + nb^{-1} \\
&\pm p_2 \pm p_3 = mb + nb^{-1},\, m,n \in \mathbb{Z}_{\geq 0} + \frac{1}{2} .
\end{align*}
This gives the wedges in $p_1 \pm p_4$ and $p_2 \pm p_3$ planes associated to degeneration when integrated tachyon is close to $p_1$. But in presence of the ground ring $p_2, p_3$ are shifted after OPE with degenerate field
\begin{align*}
&\pm p_1 \pm p_4 = mb + nb^{-1} \\
&\pm p_2 \pm p_3 + kb + lb^{-1} = mb + nb^{-1},\\
&m,n \in \mathbb{Z}_{\geq 0}  + \frac{1}{2},\,k \in \frac{1-r}{2} \dots \frac{r-1}{2},\,l \in \frac{1-s}{2} \dots \frac{s-1}{2} .
\end{align*}
It means that we have extra discrete terms for each value of $p_2 \pm p_3$, compared to $\mathsf{A}^{(b)}_{0,4}$, which can spoil the convergence.
\begin{figure}[ht]
  \centering
  \includegraphics{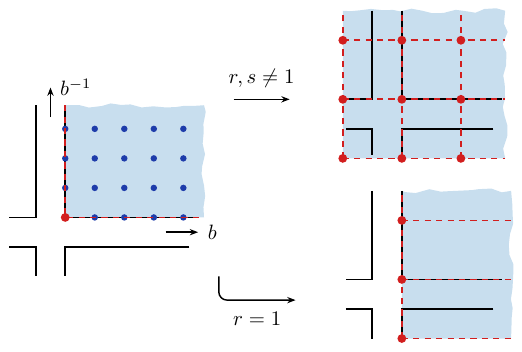}
  \caption{Analyticity properties in $p_i \pm p_j$ plane. }
  \label{quadrants}
\end{figure}

Figure \ref{quadrants} illustrates this. Left picture corresponds to the tachyon four-point amplitude. Highlighted region is the union of 90-degree wedges starting at each blue dot, where one of the contributions from discrete terms diverges and analytic continuation is ambiguous. These wedges also cover the other three quadrants on the plane (not shown on the figure); the integral over moduli space converges only in a cross-shaped region bounded by black lines. Right panels show what happens in the case of correlator with the ground ring. Because of OPE with degenerate field,  wedge associated with each discrete term (bounded by red dashed line) is copied and shifted in one (for $r=1$) or both ($r,s \neq 1$) directions in the plane by a maximum distance of $\frac{rb+sb^{-1}}{2}- \frac{Q}{2}$. In the latter case, highlighted region covers all complex plane of $p_i \pm p_j$, so the integral never converges. If $r$ ($s$) is equal to one, it still converges in the vertical (horizontal) strip. 

Note that as a function of $p_i \pm p_4$, convergence is still expected in the cross-shaped region, as for tachyon four-point amplitude. Here the wedges are not shifted, because degenerate field is never on the same side of degeneration as $p_i$ and $p_4$ together.

In the following we thus further restrict to the case $r=1$, where we still have a strip in $p_i \pm p_j$ plane where the moduli space integral converges. For general $r,s$ worldsheet integral for at least one discrete term will always diverge and one needs a separate prescription to define it (in fact, the same is true for 5-point function with degenerate tachyon $\mathcal{T}_{p_{r,s}}$).

\subsection{Values at special points}
So far we did not consider singularities of 1GR-4T correlator that appear because the integral over $z$ ceases to converge in the vicinity of ground ring insertion $x$. Such collision cannot bring any discontinuities, because this requires OPE in at least one sector to be continuous. In the following we will show that it will not bring any additional poles as well --- all potential singularities of such kind cancel for different reasons. Such analysis also allows us to compute $\mathsf{A}_{0,4|1}$ exactly at certain special values of integrated tachyon momentum $p_4$.

\paragraph{The simplest case.}
Consider the case of $O_{2,1}$. The relevant OPEs in Liouville and matter sectors read
\begin{equation}
V_{2,1}(x) V_p(z)   = (w \overline{w})^{\frac{1+b^2}{2} -  bp} C_L^+ [V_{p + b/2}(z)] + (w\overline{w})^{\frac{1+b^2}{2}+bp} C_L^- [V_{p-b/2}(z)],\label{ope}
\end{equation}
\begin{equation}
\Phi_{2,1}(x) \Phi_{ip}(z) = (w \overline{w})^{\frac{1-b^2}{2} - bp} C_M^- [\Phi_{i(p - b/2)}(z)] + (w\overline{w})^{\frac{1-b^2}{2}+bp} C_M^+ [\Phi_{i(p+b/2)}(z)],
\end{equation}
where $w=x-z$ and $C^{\pm}_{L/M}$ are degenerate structure constants, obtained from DOZZ formula. We then multiply the two OPEs, act by $H_{2,1} = \partial_x^L - \partial_x^M + b^2 \normord{\mathfrak{bc}}(x)$ and integrate the result over a small disk $|w|<\epsilon$ for fixed $x$.

Contribution of each conformal family can be written as a series in $w$.
Consider first the leading terms in $w$. If we multiply terms crisscross, we get
\begin{multline}
H_{2,1} \overline{H}_{2,1} (w \overline{w})^{\frac{1+b^2}{2} - bp} (w\overline{w})^{\frac{1-b^2}{2}+bp} C_L^+ C_M^+ U_{p+b/2}(z) = \\ = (b^2 -2 bp+ w b^2 \normord{bc}(x))\overline{(\dots)} C_L^+ C_M^+ U_{p+b/2}(x) (1+ O(|w|)).
\end{multline}
This always gives a finite contribution when integrating over the disk. On the other hand, consider ``wrong'' combinations, e.g.,
\begin{multline}
H_{2,1} \overline{H}_{2,1} (w \overline{w})^{\frac{1+b^2}{2} -  bp} (w\overline{w})^{\frac{1-b^2}{2}-bp} C_L^+ C_M^- V_{p+b/2}\Phi_{i(p-b/2)}(z) = \\  = b^4 (w\overline{w})^{-2bp} C_L^+ C_M^- V_{p+b/2}\Phi_{i(p-b/2)}(x) (1+O(|w|))
\end{multline}
We see that integral over $w$ develops a pole when $-2bp = -1;\,p = \frac{1}{2b}$, with residue independent of the radius of the disk. For other values of $p$ this contribution can be defined via analytic continuation even if it is divergent. 

For such $p$ integrated tachyon can be written as $U_{0,1} = V_{0,1} \Phi_{0,1}$. The residue of the integral at this pole is simply a correlator of remaining tachyons with an additional insertion of $V_{Q/2} \Phi_{i (b^{-1}-b)/2}$, i.e. the identity operator $O_{1,1}$.

One can also check that the same is valid when considering descendants on level $1$: when $p = \frac{1}{b}$, a pole appears with $b^8\langle O_{1,2} TTT \rangle$ as a residue (in particular, $p + \frac{b}{2} = p_{1,2}$).

One can conjecture that in general
\begin{equation}
\underset{p = \frac{k}{2b} = p_{0,k}}{\text{Res}} O_{2,1}(x) \int \limits_{|x-z|<\epsilon} d^2z\, U_p(z) = \# O_{1,k} .\label{OUresidue}
\end{equation}
If we pass to canonical normalization in CLS, we can note that exactly these values of $p$ give ``trivial zeros'' for tachyon amplitudes because of leg factors. So in fact, what we found are not the poles of the canonically normalized 1GR-4T amplitudes, but a way to compute them at these points --- zero of the leg factor cancels the pole coming from the integration near $x$ and all amplitude reduces to the residue at this pole.

It should be noted that for these values of $p$ other ingredients of OPE \eqref{ope} are singular as well. First, $C_L^-$ and $C_M^+$ have a simple pole. On the other hand, in a different conformal family (multiplied by $C_L^+$ and $C_M^-$) poles appear in conformal block expansion coefficients (e.g., for $p \to \frac{1}{2b}$ on level $1$), since intermediate field becomes degenerate. We do not expect poles for each Liouville correlator separately at this values of $p$, and, in fact, two sources of poles that we mention cancel each other. The mechanism is based on HEM: for example, contributions from conformal family of $V_{p-\frac{b}{2}} \to V_{-1,1}$, multiplied by divergent structure constant, are cancelled by descendants of $L_{-1} \overline{L}_{-1} V_{p+\frac{b}{2}} \to L_{-1} \overline{L}_{-1} V'_{1,1}$, which come with divergent coefficients from conformal blocks. This is described in more detail in \cite{Khromov:2025awh}.

We can compute the normalization coefficient in \eqref{OUresidue}. It comes from the product of structure constants, coefficient coming from the action of $H_{2,1}$, leg factor and the result of integration over $d^2w$ 
\begin{equation}
C_L^+ C_M^-(p =\frac{k}{2b})\underset{p \to \frac{k}{2b}}{\lim}  \left(\mathcal{N}(p)\cdot \frac{2\pi }{2 (k-2 b p)} \right) \cdot 
\begin{cases}
b^4,k=1 \\
b^8, k=2 \\
\frac{b^{12}}{16},k=3 \\
\dots
\end{cases} .\label{normf}
\end{equation}
The last coefficient comes from the action of $H_{2,1}$ on descendants on the corresponding level (we computed it explicitly for the listed cases). Canonical normalization for ground ring operators is $\mathcal{O}_{m,n} = -\frac{\pi \mathcal{N}(p_{m,-n})}{B_{m,n}} O_{m,n}$. So \eqref{normf} should additionally be multiplied by $\frac{\mathcal{N}(p_{2,-1})}{\mathcal{N}(p_{1,-k})} \frac{B_{1,k}}{B_{2,1}}$. Collecting all together, we obtain a very simple answer
\begin{equation}
\lim\limits_{p \to \frac{k}{2b}} \mathcal{O}_{2,1}(x) \int \limits_{|z-x|<\epsilon} d^2z\,\mathcal{U}_p(z) = -b\mathcal{O}_{1,k} .
\end{equation}
Note in particular that this is independent of $k$. We checked that this is true for $k=1,2,3$, where explicit expressions for $H_{1,k}$ are available. 

\paragraph{Case of $O_{3,1}$.} Consider now the OPE
\begin{align*}
&V_{3,1}(x) V_p(z) = &\overset{(1)}{C_L^+ (w\overline{w})^{1+b^2 - 2bp} V_{p+b}} + \quad&\overset{(2)}{C_L^0 (w\overline{w})^{1+2b^2}V_{p}}+ \quad &\overset{(3)}{C_L^- (w\overline{w})^{1+b^2+2bp}V_{p-b}}, \\
&\Phi_{3,1}(x) \Phi_{ip}(z) = &C_M^- (w \overline{w})^{1-b^2 - 2bp} \Phi_{i(p-b)}+ \quad &C_M^0 (w\overline{w})^{1-2b^2}\Phi_{ip}+  &C_M^+ (w\overline{w})^{1-b^2+2bp}\Phi_{i(p+b)} .\\
\end{align*}
We act on this by operators $H_{3,1} \overline{H}_{3,1}$ which reduces powers of $w$ at most by $2$. Again we look for poles coming from $d^2w$ integration, first restricting to the primary field level. No poles would appear if we multiply contributions $(1)(3), (2)(2)$ or $(3)(1)$: in this combinations Liouville and matter dimensions add up to $1$.  Consider other options:

\begin{itemize}
    \item $(1)(2)$: leading powers are
    $
    (w\overline{w})^{2-b^2-2bp}
    $; possible pole after acting with $H_{3,1}$ is when $$-b^2 - 2bp = -1; p = \frac{1}{2b} - \frac{b}{2} = p_{-1,1}.$$
    In this case $U_p$ is dressed matter unit operator. This means that the OPE $\Phi_{3,1} \Phi_{ip}$ should truncate to the first term and $C_M^0$ goes to zero, cancelling this pole
    \item $(2)(1)$: leading powers are
    $
    (w\overline{w})^{2+b^2-2bp}
    $; possible pole after acting with $H_{3,1}$ is when $$ -2bp = -1 - b^2; p = \frac{1}{2b} + \frac{b}{2} = p_{1,1}.$$ Now we perform OPE with the dressed Liouville unit operator; everything is the same as before, but the pole is cancelled by the zero of $C_L^0$.
\end{itemize}
For both these values of $p$ there is no zero in the leg factor; although the pole cancels, there are other finite contributions in the OPE and we cannot use the previous arguments to reduce correlator to the residue. However, these poles will still be important in what follows, see section \ref{subs-hem-fixed}. Cases $(3)(2)$ and $(2)(3)$ are symmetric up to the choice of sign of $p$.

The remaining case is $(1)(1)$ (and the symmetric one $(3)(3)$). Here the pole condition for primary field contributions would read
$p = \frac{1}{4b}$. One can check that this pole is cancelled by the factors that appear when acting with $H_{3,1}$, which give zero precisely for $p = \frac{1}{4b}$. 

If in $(1)(1)$ term we consider subleading powers of $w \overline{w}$, the condition for the pole takes the form $1 - 4bp = -1; p = \frac{1}{2b}$. This can be cancelled by the leg factor, familiar from previous example; again for this value of $p$ the correlator is determined only by this term. It is proportional to 1GR-3T correlator with $\mathcal{O}_{2,1}$ and the remaining 3 tachyons, with the coefficient
$$
\frac{\mathcal{N}(p_{-3,1})}{\mathcal{N}(p_{-2,1})} \frac{B_{2,1}}{B_{3,1}}C_L^+ C_M^-(p =\frac{1}{2b})\underset{p \to \frac{1}{2b}}{\lim}  \left(\mathcal{N}(p) \cdot \frac{2\pi  }{2 (2-4 b p)}\right) \cdot \left(4b^2\right)^2.
$$
The last factor comes from the action of $H_{3,1}$. Collecting all the factors together, we can get 
\begin{equation}
\lim\limits_{p \to \frac{1}{2b}} \mathcal{O}_{3,1}(x) \int \limits_{|z-x|<\epsilon} d^2z\,\mathcal{U}_p(z) = -2b\mathcal{O}_{2,1} .
\end{equation}
The total coefficient is $(-2b)$ -- twice as for $\mathcal{O}_{2,1}$. 

This makes sense in view of the ground ring algebra: we can replace $\mathcal{O}_{3,1}$ with two generators of the ring $\mathcal{O}_{2,1}(x+\epsilon) \mathcal{O}_{2,1}(x-\epsilon)$, separated by some distance. On the other hand, when $p \to p_{0,1}$ the correlator reduces to two contributions in the vicinity of $x \pm \epsilon$:
\begin{equation}
\langle \mathcal{O}_{2,1}(x+\epsilon) \mathcal{O}_{2,1}(x-\epsilon) \int d^2z\, \mathcal{U}_p(z)  \dots \rangle \mid_{p=p_{0,1}}= \sum \limits_{\pm}\left(-b\right) \langle \mathcal{O}_{2,1}(x\pm\epsilon) \dots \rangle = - 2b\langle \mathcal{O}_{2,1}(x) \dots \rangle.
\end{equation}
Alternatively, perform the OPE $\mathcal{O}_{2,1}(x+\epsilon) \mathcal{O}_{2,1}(x-\epsilon)  = \mathcal{O}_{3,1}(x) + \mathcal{O}_{1,1}(x) + \dots $. The second term will not contribute for such values of $p$ and the first gives the same answer.

The upshot is that integrated tachyon (at least for these special values of $p$) does not interfere with the ground ring algebra. It does, however, interfere with $\mathcal{O} \mathcal{T}$-OPE, as we discussed before. 

\paragraph{General case. } One can use this observation to conjecture the values for arbitrary $\mathcal{O}_{r,s}$ insertion at points $p=p_{0,k}$, presenting $\mathcal{O}_{r,s}$ as a Chebyshev polynomial in generators $\mathcal{O}_{1,2}$, $\mathcal{O}_{2,1}$. Assuming that we can use the ground ring algebra: as mentioned before, general GR operator can be written as 
\begin{equation}\mathcal{O}_{r,s} = \mathcal{O}_{r,1} \mathcal{O}_{1,s} =
\mathbf{U}_{r-1} (\mathcal{O}_{2,1}/2) \mathbf{U}_{s-1} (\mathcal{O}_{1,2}/2),
\end{equation}
where $\mathbf{U}$ is the Chebyshev polynomial. Each term is understood in the point-split sense. When $p \to \frac{k}{2b}$, each term reduces to contributions from the neighborhoods of $\mathcal{O}_{2,1}$ insertions. More precisely, 
\begin{equation}
\langle \mathcal{O}_{1,s}  \left( \frac{\mathcal{O}_{2,1}}{2} \right)^l \int U_p(z)\, d^2z \dots \rangle \mid_{p = \frac{k}{2b}} = -\frac{lb}{2} \langle  \mathcal{O}_{1,s} \mathcal{O}_{1,k} \left( \frac{\mathcal{O}_{2,1}}{2} \right)^{l-1}  \dots \rangle.
\end{equation}
This means that
\begin{equation}
\langle \mathcal{O}_{1,s}  \mathbf{U}_{r-1} (\mathcal{O}_{2,1}/2) \int U_p(z)\, d^2z \dots \rangle \mid_{p = \frac{k}{2b}} =- b \langle  \mathcal{O}_{1,s} \mathcal{O}_{1,k} \cdot \frac{1}{2}\mathbf{U}'_{r-1} (\mathcal{O}_{2,1}/2)  \dots \rangle.
\end{equation}
$\mathbf{U}$ form a basis, so $\mathbf{U}'_{r-1}$ can again be expanded in ground ring operators: the result is
\begin{equation}
\frac{1}{2} \mathbf{U}'_{r-1} (x) = \sum \limits_{0 \leq l \leq r-1,  l = r-1 \text{ mod }2} l\, \mathbf{U}_{l-1}(x),
\end{equation}
so we have
\begin{equation}
 b \langle  \mathcal{O}_{1,s} \mathcal{O}_{1,k} \cdot \frac{1}{2}\mathbf{U}'_{r-1} (\mathcal{O}_{2,1}/2)  \dots \rangle =  b \sum \limits_{0 \leq l \leq r-1,  l = r-1 \text{ mod }2}  l\langle  \mathcal{O}_{1,s} \mathcal{O}_{1,k} \mathcal{O}_{l,1}  \dots \rangle.
\end{equation}
For the correlator we consider, this in turn reduces to a tachyon three-point function (perhaps with one of the arguments shifted by half-periods $b/2, b^{-1}/2$, depending on the parity of $r,s,k$)
\begin{multline}
    -b \sum \limits_{l}  l\langle  \mathcal{O}_{1,s} \mathcal{O}_{1,k} \mathcal{O}_{l,1}  \mathcal{T}_{p_1} \mathcal{T}_{p_2} \mathcal{T}_{p_3} \rangle = -b \cdot ks \langle  \mathcal{T}_{p_1 + b \frac{r}{2} + \frac{(k+s)}{2b} } \mathcal{T}_{p_2} \mathcal{T}_{p_3}\rangle\cdot  \sum \limits_{l} l^2 = \\=- b  \frac{ks\,r(r^2-1)}{6} \mathsf{A}_{0,3}^{(b)}(p_1+ \frac{kb^{-1}}{2}-\frac{Q}{2}, p_2, p_3)
\end{multline}
(in the last line we assumed that $r,s$ are odd; we also used periodicity of tachyon 3-point function and the fact that it does not matter which of the momenta to shift).

There is, of course, an analogous expression for 1GR-4T correlator at the point $p = \frac{k b}{2}$, obtained by exchanging the roles of $r$ and $s$ and $b \to b^{-1}$.
\subsection{Bootstrapping $\delta \mathsf{A}$}
Let us collect together the constraints that we obtained for a meromorphic correction $\delta \mathsf{A}(p_4,\vec{p}|1,s)$:
\begin{enumerate}
    \item when parameters of stationary tachyons $p_i = \frac{l b^{-1}}{2},\,l \in \mathbb{Z}$, $i = 1,2,3$, we expect that the worldsheet integral converges and the amplitude vanishes after multiplication by leg factors. Since $\mathsf{A}_{0,4} (\boldsymbol{p}) = 0$ at this point, $\delta \mathsf{A} =0$ as well. We cannot state the same about $p_i = \frac{l b}{2}$, since in this case worldsheet integral diverges and the integral requires analytic continuation under the tachyon cuts.
    \item when parameter of integrated tachyon $p_4 = \frac{lb^{-1}}{2}$, we also get zero. For $p_4 = \frac{lb}{2}$, from the results of the previous paragraph we should get
    \begin{equation}
    \delta \mathsf{A}(\frac{lb}{2},\vec{p}|1,s) = - l\frac{s(s^2-1) }{6b} \mathsf{A}_{0,3}^{(b)}(p_1+ \frac{lb}{2}-\frac{Q}{2},p_2, p_3).   \label{deltaA} 
    \end{equation}
    \item $\delta \mathsf{A}$ can have poles only at $\pm p_1 \pm p_2 \pm p_3 \pm p_4 \in \ZZ b+\ZZ b^{-1}$.
\end{enumerate}
Additionally $\delta \mathsf{A}$ must be odd in all momenta. Consider then the general form of meromorphic function with appropriate properties:
\begin{equation}
    \sum_{m,n \in \mathbb{Z}} \sum_{\sigma_1,\dots,\sigma_4=\pm} \sigma_1 \sigma_2\sigma_3\sigma_4 \frac{f(\sigma_1 p_1,\sigma_2 p_2,\sigma_3p_3,\sigma_4p_4,m,n)}{\sum_{j=1}^4\sigma_j p_j+m b+n b^{-1}}\ .
\end{equation}
A priori $f$ is symmetric only in permutations of stationary tachyons. We can assume that residues do not grow too fast, i.e., $f$ is polynomial.  Under flipping all signs of $m$ and $n$ and all the $\sigma_i$, even degree part of $f$ cancels. Among odd degree polynomials the ones that are divisible by the denominator also cancel after summation over signs $\sigma_i$; so in total we consider odd degree, symmetric in $p_{1,2,3}$ polynomials modulo the ideal generated by the denominator.

One can show that there is a unique solution to above constraints (which must hold for arbitrary values of other parameters, which is quite restrictive) assuming $f$ is a polynomial of degree $\leq 3$. It reads
\begin{equation}
f = i\frac{s(s^2-1)}{48\pi b}m.
\end{equation}
So, in fact, we could have just restricted to linear functions. It is unlikely that new nontrivial solutions would appear for $f$ of higher degree.

Let us check that it reduces to $\mathsf{A}_{0,3}$ at special points. The three-point amplitude can be represented as \cite{Collier:2024kwt}
\be 
\mathsf{A}_{0,3}^{(b)}(\boldsymbol{p})=\frac{i}{8\pi} \sum_{m,n \in \ZZ+\frac{1}{2}}\sum_{\sigma_1,\sigma_2,\sigma_3=\pm} \frac{\sigma_1\sigma_2\sigma_3}{\sigma_1 p_1+\sigma_2 p_2+\sigma_3 p_3+m b+n b^{-1}}\ .
\ee
Then one has, for $p_4 = \frac{lb}{2}$
\begin{align}
    &\sum_{m,n \in \mathbb{Z}} \sum_{\sigma_1,\dots,\sigma_4=\pm}  \frac{\sigma_1 \sigma_2\sigma_3\sigma_4 m}{\sum_{j=1}^4\sigma_j p_j+m b+n b^{-1}}=\nonumber \\=& \sum_{m,n \in \ZZ} \sum_{\sigma_1,\sigma_2,\sigma_3=\pm}\bigg[ \frac{\sigma_1\sigma_2\sigma_3m}{\sum_{i=1}^3 \sigma_i p_i+(m+\frac{l}{2})b+nb^{-1}}- \frac{\sigma_1\sigma_2\sigma_3 m}{\sum_{i=1}^3 \sigma_i p_i+(m-\frac{l}{2})b+nb^{-1}}\bigg] = \nonumber \\=&\sum_{m \in \ZZ+\frac{l}{2},n \in \ZZ} \bigg[\sum_{\sigma_1,\sigma_2,\sigma_3=\pm} \frac{\sigma_1\sigma_2\sigma_3(m-\frac{l}{2})}{\sum_{i=1}^3 \sigma_i p_i+mb+nb^{-1}}-\sum_{\sigma_1,\sigma_2,\sigma_3=\pm} \frac{\sigma_1\sigma_2\sigma_3(m+\frac{l}{2})}{\sum_{i=1}^3 \sigma_i p_i+mb+nb^{-1}}\bigg] = \nonumber\\=& 8\pi i l\, \mathsf{A}_{0,3}^{(b)}(p_1,p_2,p_3+\frac{lb}{2}- \frac{Q}{2})\ .
\end{align}

Up to coefficient the sum over poles corresponding to such choice of $f$ is actually the same as the function $\mathsf{C}$, defined previously. Thus, bootstrap considerations lead us to the following answer
\begin{equation}
\mathsf{A}_{0,4|1}(p_4,\vec{p}|(1,s)) = s\mathsf{A}^{(b)}_{0,4}(p_4,\vec{p}) -\frac{ s(s^2-1)}{3} \mathsf{C}. \label{fixedGRperiod}
\end{equation}
It is somewhat surprising that in the end the answer turns out to be completely symmetric in all $4$ tachyon parameters. We will see later that this is a peculiarity of this case and for higher-point amplitudes integrated tachyons play a distinguished role.

Note in particular that $\mathsf{A}_{0,4|1}$ differs from a sum of shifted four-point functions, that one would obtain from naive $\mathcal{O} \mathcal{T}$-OPE arguments
\begin{equation}
\mathsf{T}_{0,4|1}(\boldsymbol{p}|(1,s)) =\sum_{\ell = (1-s)/2}^{(s-1)/2}  \mathsf{A}_{0,4}^{(b)}(\boldsymbol{p} \setminus p_i,p_i-b^{-1} \ell) = s\mathsf{A}_{0,4}^{(b)}(\boldsymbol{p}) - \frac{s(s^2-1)}{6} \mathsf{C}(\boldsymbol{p}). \label{sumofs}
\end{equation}
However, $\mathsf{T}_{0,4|1}$ is also symmetric with respect to all permutations of $p_1, \dots, p_4$.
\subsection{HEM and their modification} \label{subs-hem-fixed}
There is another set of constraints on $\delta \mathsf{A}$ that one might want to use. Namely, if the integrated tachyon is degenerate (in one of the sectors, for concreteness, let us pick $p_4 = \frac{mb-n/b}{2}$), one can use HEM to reduce the $z$-integral to boundary contributions. As in the case of $4$-point function, the integral can be expressed as a sum of ``contact terms'' when $z \to x$ or $z_i$ and ``contribution from infinity''. The last two options together bring
\begin{multline}
\sum \limits_{i=1}^3\sum \limits_{k=-(m-1):2}^{m-1} \sum \limits_{l=-(n-1):2}^{n-1}  q_{k,l}^{m,n} (p_i) \langle \mathcal{O}_{1,s} \mathcal{T}_{p_i + \frac{kb}{2} + \frac{lb^{-1}}{2}} \dots \rangle - 2 p_{m,n} \langle\mathcal{O}_{m,n} \mathcal{O}_{1,s} \mathcal{T}_{p_1}  \mathcal{T}_{p_2} \mathcal{T}_{p_3}  \rangle   =\\ = s\left(mn p_{m,n} - \sum \limits_{i=1}^3\sum \limits_{k=-(m-1):2}^{m-1} \sum \limits_{l=-(n-1):2}^{n-1}\left|p_i - \frac{k b}{2} - \frac{l b^{-1}}{2}\right| \right) \mathsf{A}_{0,3} (p_1 + p_{m,n}- \frac{Q}{2},p_2,p_3)\label{firstcont}.
\end{multline}
For odd $s$, since three-point function is doubly periodic, we can ignore the shifts of the arguments caused by $\mathcal{O}_{1,s}$; then we have the answer in the last line --- $s$ times the tachyon four-point function, computed using HEM. Hence the boundary contribution when $z \to x$ should agree with $\delta \mathsf{A}$.

This requires the logarithmic terms in the $\mathcal{O}'_{m,n} \mathcal{O}_{1,s}$ OPE. As for the tachyon case, they appear because we differentiate the momentum-dependent coordinate factor. That means that the coefficients are $q$-factors, evaluated at degenerate Liouville momenta:
$$
\mathcal{O}'_{m,n}(z) \mathcal{O}_{1,s}(x) = \log|z-x|^2 \sum \limits_{\Phi_{k,l} \in [\Phi_{m,n}][\Phi_{1,s}]} \left(\frac{mb+nb^{-1}}{2}- \left|\frac{kb+lb^{-1}}{2}\right| \right)\mathcal{O}_{k,l}(x) + \dots
$$
The sum is over such $(k,l)$ that $\Phi_{k,l}$ appears in the fusion $\Phi_{m,n} \Phi_{1,s}$. Using this OPE, ``contact term'' when $z \to x$ is then
\begin{equation}
    \sum \limits_{(k,l)} kl \left(\frac{mb+nb^{-1}}{2}- \left|\frac{kb+lb^{-1}}{2}\right| \right) \mathsf{A}_{0,3} (p_1 + p_{m,n}- \frac{Q}{2},p_2,p_3).
\end{equation}
The sum over $k,l$ evaluates to
\begin{equation}
 \sum \limits_{(k,l)} kl \left(\frac{mb+nb^{-1}}{2}- \left|\frac{kb+lb^{-1}}{2}\right| \right) = 
 \begin{cases}
    - \frac{ms(s^2-1)}{6b},\,n \geq s \\
    -\frac{m n \left(n^2-3 n s+3 s^2-1\right)}{6 b}, n < s \\
 \end{cases} \label{deltaAhem}.
\end{equation}
Since we have
\begin{align*}
&\mathsf{C}(\frac{lb+ nb^{-1}}{2},p_1 \dots p_3 ) = \sum_{m=1}^\infty (-1)^{mn}\frac{\prod_{i=1}^3\sin(2\pi m b p_i)}{\sin(\pi m b^2)} \frac{  \sin (\pi l m b^2)}{\sin(\pi m b^2)} = \\
&= \sum_{m=1}^\infty (-1)^{mn}\sum \limits_{k=(1-l):2}^{l-1}\frac{\prod_{i=1}^3\sin(2\pi m b p_i)}{\sin(\pi m b^2)}  \cos \pi k mb^2 = \\
& =\sum_{m=1}^\infty (-1)^{m}\sum \limits_{k=(1-l):2}^{l-1} \sum \limits_{\pm} \frac{1}{2} \frac{\sin (2\pi m b(p_1+ \frac{(n-1)b^{-1}}{2} \pm \frac{kb}{2}))\prod_{i=2}^3\sin(2\pi m b p_i)}{\sin(\pi m b^2)} = \\
& = \frac{l}{2b} \mathsf{A}_{0,3}^{(b)}(p_1+ \frac{lb+nb^{-1}}{2}-\frac{Q}{2},p_2, p_3),
\end{align*}
the $-\frac{s(s^2-1)}{3} \mathsf{C}$ term in our ansatz agrees with HEM result only in the case $n \geq s$.

The resolution of this apparent disagreement is in the previously discussed additional singularities in $\mathcal{O} \mathcal{U}$-OPE. The model example is the collision $\mathcal{O}_{1,3} \int d^2z\, \mathcal{U}_{1,-1}$: recall from the previous discussion that we found divergence in the $z$-integral, which is compensated by vanishing structure constant $C^0$ in one of the sectors. This is missed by naive HEM calculation above which assumes truncated degenerate fusion. For this example, difference between two options in \eqref{deltaAhem} is equal to $b^{-1}$; taking care of all normalization factors, this is precisely reproduced by a finite limit of this ``forbidden'' contribution to the OPE, integrated over the vicinity of $x$.

\subsection{Periods and relations in homology of $\overline{\mathcal{M}}_{0,5}$} \label{subs-periods}
Independently, one can look for an expression for another period of the GR 2-form, namely,
\begin{equation}
\mathsf{S}_{0,4|1}(\boldsymbol{p}|(1,s)) = - \frac{\pi \mathcal{N}(\frac{b}{2}-\frac{s}{2b})}{B_{1,s}}\int d^2x\, \langle  R_{1,s} \overline{R}_{1,s} V_{1,s}\Phi_{1,s}(x)  \mathcal{T}_{p_1}(y) \mathcal{T}_{p_{2}}(0) \mathcal{T}_{p_{3}}(1) \mathcal{T}_{p_{4}}(\infty)   \rangle.
\end{equation}
This expression is manifestly symmetric in $p_1, \dots p_4$, does not have any discontinuities (they cannot appear from the collision of degenerate field and tachyon) and is thus meromorphic, but the values at special points $p_i = \frac{ l b^{\pm 1}}{2}$ can be computed in exactly the same way as for $\mathsf{A}_{0,4|1}$. In fact, at these points $\mathsf{S}$ coincides with $\delta \mathsf{A}$ given in \eqref{deltaA}. Repeating the previous bootstrap argument brings us to the answer
\begin{equation}
 \mathsf{S}_{0,4|1}(\boldsymbol{p}|(1,s)) =  - \frac{s(s^2-1)}{3} \mathsf{C}(\boldsymbol{p}). \label{fixtachyonperiod}
\end{equation}
A nice sanity check is that this can be reproduced in a different way. Note that the sum of four-point functions with shifted momenta $\mathsf{T}_{0,4|1}$ can be naturally interpreted as an integral of GR 2-form $\Omega_{1,s}$ over the boundary divisor $\Delta_{x\,i}$ in $\overline{\mathcal{M}}_{0,5}$ where $x$ and $z_i$ collide ($i=1,2,3$, but the result is independent of $i$). Indeed, on such strata we can trust $\mathcal{O} \mathcal{T}$-OPE for the tachyon which is on the same ``bubble'' as GR operator. The dimension of second (co)homology group of $\overline{\mathcal{M}}_{0,5}$ is five and we already know several periods of the GR 2-form; we can check whether these periods conform with the relations in homology.

Let $F_{x,y}$ be 2-cycles which we integrate over when keeping constant $y$ and $x$ respectively (i.e., the fibers of the corresponding forgetful maps); then we have in homology
\begin{equation}
F_x = \Delta_{y1} + \Delta_{23},\,F_y = \Delta_{x1} + \Delta_{23}.
\end{equation}
This relation is obvious if we consider a fiber of forgetful map over the degenerate curve, e.g., $(y1|23) \in \partial\overline{\mathcal{M}}_{0,4}$. For the smooth point it then follows because the forgetful map is a ``flat morphism''. In fact we can reorder $4$ points $(y,1,2,3)$ in the RHS in an arbitrary way and we have the same homology class, which gives a particular case of Keel relations \cite{Keel:1992int} in $\overline{\mathcal{M}}_{0,n}$.

$\Delta_{y1}, \Delta_{23}$ are divisors where two tachyons collide; given the symmetry of other quantities, we may assume the integrals of GR 2-form over these two divisors are equal as well. This predicts 
\begin{equation}
\int \limits_{\Delta_{y1}} \Omega_{1,s} = - \frac{s(s^2-1)}{6} \mathsf{C}(\boldsymbol{p}) .\label{gr2formotherdivisor}
\end{equation}
We will show how this can be computed later in section \ref{sec-codim-2}. Then we must have
\begin{equation}
0 \overset{?}{=} \left[\int \limits_{F_x} - 2 \left(\int \limits_{F_y} - \int \limits_{\Delta_{x1}}\right)\right] \Omega_{1,s} = \mathsf{S}_{0,4|1}(\boldsymbol{p}|(1,s)) - 2 \mathsf{A}_{0,4|1}(\boldsymbol{p}|(1,s))+2 \mathsf{T}_{0,4|1}(\boldsymbol{p}|(1,s)).
\end{equation}
This is indeed correct for our expressions.

\section{Higher-point tachyon correlators and HEM} \label{sec: higherptHEM}
 Now we will use the results of the previous section  to study degenerate five-point tachyon amplitude \eqref{5pfdeg}
\begin{equation}
 C_{S_2}^{(b)} \int d^2x\,d^2y \left\langle \mathcal{U}_{\frac{b+sb^{-1}}{2}}(x) \dots \mathcal{U}_{p_2}(y) \mathcal{T}_{p_{1}}(0) \mathcal{T}_{p_{3}}(1) \mathcal{T}_{p_{4}}(\infty) \right\rangle,
\end{equation}
 using higher equations of motion. We mentioned above that accounting only for contributions known from the four-point calculation reproduces only a part of the ``matrix model'' result. We will explain now where the extra contributions come from and reproduce the full answer \eqref{5pfdeg} from HEM.

\subsection{Explaining the curvature term}
The curvature term in tachyon five-point HEM should match the first term in \eqref{5pfdeg}, proportional to $\frac{s-b^2}{b}$. This fails if we only consider the $\mathsf{A}_{0,4|1}$-period, as we did for four-point function.

The correct way to compute the curvature term now is as follows: singularities of $\Omega'_{1,s}$ appear on the strata in $\overline{\mathcal{M}}_{0,5}$ where $x$ becomes a bad coordinate, i.e., where $dx$, considered as a section of cotangent line bundle $\mathbb{L}_x$ at marked point $x$, has zeros or poles. This means we need to transform to a different coordinate chart, which produces logarithms due to inhomogeneous transformation law of $O'$. The logarithmic terms we obtain are proportional to the form $\Omega$ on these strata; together they combine into the curvature term. Coefficient with which each boundary divisor enters is determined by the degree of zero or pole that the $1$-form $dx$ has at each component, since it determines what change of coordinates we should use. Bad strata include $7$ codimension $1$ degenerations --- every irreducible boundary component apart from the ones where $y$ collides with $z_i$ (zeros of order $1$), as well as $x=\infty$ (pole of order $2$; the corresponding divisor is the same as the fiber of forgetful map $F_y$). The total divisor of $dx$ can be written as follows
\begin{equation}
D(dx) = \sum \limits_{i=1}^3 \Delta_{xi} + \Delta_{xy} - 2 F_y + \sum \limits_{1 \leq i < j \leq 3} \Delta_{ij} \label{div-dx}\ .
\end{equation}
Taking into account that $\int \limits_{\Delta_{ij}} \Omega_{1,s} = \frac{1}{2}\int \limits_{F_x}\Omega_{1,s}$ (see \eqref{gr2formotherdivisor} above), if we sum all the contributions, we get
\begin{equation}
4\mathsf{T}_{0,4|1}(\boldsymbol{p}|(1,s)) - 2\mathsf{A}_{0,4|1}(\boldsymbol{p}|(1,s)) + \frac{3}{2}\mathsf{S}_{0,4|1}(\boldsymbol{p}|(1,s)) = 2 (s\mathsf{A}_{0,4}^{(b)}(\boldsymbol{p}) - \frac{s(s^2-1)}{4} \mathsf{C}(\boldsymbol{p})).
\end{equation}
The relative coefficient between $\mathsf{A}_{0,4}$ and $\mathsf{C}$ matches the first term in \eqref{5pfdeg} precisely. 

A short calculation shows that the cohomology class dual to $D(dx)$ is, in fact, just the usual $\psi$-class associated to point $x$. So, curvature term computes the ``first descendant'' of the ground ring 2-form, or intersection number with the $\psi$-class. We expect that this is a general answer. 

Note that if we take the combination of tachyon five-point functions with degenerate fields in two sectors, i.e.,
$
\mathsf{A}_{0,5}(p_{1,s}) + \mathsf{A}_{0,5}(p_{-1,s})
$, 
everything but the curvature term cancels. This clarifies a proposal of \cite{Artemev:2025pvk} for  minimal string, where a similar linear combination was suggested to compute ``ground ring correlators''. In minimal string it is less trivial to interpret the first term, since tachyon operator with degenerate field in Liouville sector does not exist; so, unlike CLS, it is a new element of the dictionary, not expressed in terms of already existing tachyon correlators.

\subsection{The remainder from ``codimension-2'' degeneration} \label{sec-codim-2}
The thing that is left now is to interpret  the remainder $\mathsf{R} =\frac{(s^2-1)^2}{16b} \mathsf{C}(\boldsymbol{p}) - \frac{s^2-1}{2b} \mathsf{E}(\boldsymbol{p})$. It does not have ``absolute value'' factors $|p_i-l/b|$ nor is it proportional to $p_{-1,s}$, so it cannot be explained by codimension-1 singularities at boundary divisors coming from logarithmic terms in HEM. So the remainder must come from a separate region of the moduli space.

The answer where to look for it was actually given in the work of Polchinski \cite{Polchinski:1990ia}. To our knowledge this is the first attempt to use higher equations of motion to compute minimal string amplitudes (although for the simplest case of pure gravity, or $(2,3)$ minimal string, and in modified form), the idea that was followed up on in \cite{Zamolodchikov:2003yb} and \cite{Belavin:2006ex}. He identified certain finite contributions to the correlator that come from ``corners'' of the moduli space, i.e., codimension-2 degenerations, when two pinches appear that separate ground ring operator from the rest of the surface. In pure gravity this makes HEM recursion on level $2$ consistent with known results for amplitudes in genus zero. Away from pure gravity case in $(2,2\mathbf{p}+1)$ minimal string, as we will comment on later in \ref{sec-minimalstringfail}, some additional subtleties appear that do not allow for clean generalization of Polchinski's argument. The CLS case, however, turns out to be treatable in a very similar way.

We study the regularized integral
$\left(-\frac{1}{\pi}
\int \limits_{1>|x|>|y|>\epsilon} \partial \overline{\partial} \Omega_{1,s}^{\text{log}}\right)
$. The contribution relevant in what follows appears from the boundary term at $|y| = \epsilon$, coming from integration by parts of $\partial_y \overline{\partial}_y \Omega^{\text{log}}_{x\overline{x}}$. In the following we will consider only ``spinless'', i.e., dependent only on $|x|$ and $|y|$, contributions to this 2-form (only they can be finite and cutoff-independent).

Introduce $$\log \frac{1}{\epsilon^2} = t,\,\frac{\log \frac{1}{|x|^2}}{t} = u. $$
In the limit $t \to \infty$ with finite $u \in(0,1)$ we study the vicinity of the moduli space corner. Double degeneration happens in this limit since $1 \gg|x| \gg |y|$, with $y$ and $0$ bubbling off and $x$ separated from the other points by $2$ nodes. Integrating over polar angle (replacing $d^2x = \pi |x|^2 t\, du$), this contribution is reduced to
\begin{equation}
-\pi \int \limits_0^1 du\,|x|^2\Omega_{x\overline{x}}^{\text{log}} .
\end{equation}
Thus we need to understand how $\Omega_{x\overline{x}}^{\text{log}}$ behaves in the considered limit. 

\paragraph{Saddle point calculation at the boundary.} We start with a simpler question and determine the behaviour of $\Omega_{y\overline{y}}$ (without the log) in this limit. The integrand in this case is a correlator with the fixed ground ring, which we compute using the resolution of identity (restricting it to the level of primary fields and lowest ghost excitations):
\begin{multline}
\langle \mathcal{T}_4(\infty) \mathcal{T}_3(1) \mathcal{O}_{1,s}(x) \mathcal{U}_{2}(y) \mathcal{T}_1(0) \rangle =\\ = \frac{1}{4} \int dp\,dp_M \rho_b(p) \rho_{-ib}(p_M) \langle \mathcal{T}_4(\infty) \mathcal{T}_3(1) \mathcal{O}_{1,s}(x) \mathfrak{c\overline{c}} V_p \Phi_{p_M}(0)\rangle \times \\
\times \underbrace{\langle \mathfrak{c}_{-1} \overline{\mathfrak{c}}_{-1} \mathfrak{c}_{0} \overline{\mathfrak{c}}_0 \mathfrak{c}_{1} \overline{\mathfrak{c}}_1  \rangle_{gh}}_{=1} |y \overline{y}|^{-p^2 - p_M^2-1} \langle V_p \Phi_{p_M}(\infty) \mathcal{U}_{P_2}(1) \mathcal{U}_{P_1}(0) \rangle+ \dots
\end{multline}
We assume that $\mathcal{T}, \mathcal{U}, \mathcal{O}$ are already canonically normalized. OPE of ground ring operator with the intermediate state has ``diagonal'' (when $p$ and $p_M$ are shifted in the same way) and ``non-diagonal'' terms. Non-diagonal terms are vanishing when intermediate state goes on shell $p = i p_M$, so we can write
\begin{multline}
\mathcal{O}_{1,s}(x) \mathfrak{c\overline{c}} V_p \Phi_{p_M}(0) = \sum \limits_{k = \frac{1-s}{2}}^{\frac{s-1}{2}} \mathcal{A}_k(p,p_M) (x\overline{x})^{-\frac{2k}{b}(p-ip_M)} \mathfrak{c\overline{c}}  V_{p+ \frac{k}{b}} \Phi_{p_M- \frac{ik}{b}}(0) + \\+ O(p-ip_M) + \text{descendants.}
\end{multline}
In canonical normalization OPE coefficients with on-shell tachyon are equal to $1$, so
\begin{equation}
\mathcal{A}_k(p,-ip) = \frac{\mathcal{N}(p + \frac{k}{b})}{\mathcal{N}(p)}.
\end{equation}
Now the Polchinski argument --- for $x,y \to 0$ power factors allow to compute $(p,p_M)$ integral by saddle point. Denoting as before $|y|^2 = e^{-t}, |x|^2 = e^{-ut},t \to \infty$ for the considered terms we have the momentum-dependent exponential
\begin{equation}
e^{t(p^2 + p_M^2 + \frac{2 k u}{b}(p-ip_M))} \label{gauss-int}
\end{equation}
The saddle point is then $p = ip_M = - \frac{ku}{b}$.  Intermediate state is forced to be on-shell, and the expression in the exponential vanishes. Note that $pb \in \mathbb{R}$, so the saddle point lies on the integration contour. Higher descendants and ``non-diagonal'' terms, which we do not consider, will be suppressed either in $1/t$ or exponentially. For these leading contributions to $\Omega_{y\overline{y}}$ we have, in the end
\begin{equation}
\Omega_{y\overline{y}} = \frac{2\pi}{4t} \sum \limits_{k= \frac{1-s}{2}}^{\frac{s-1}{2}} \frac{1}{y\overline{y}} \frac{\rho_b(-\frac{ku}{b}) \rho_{-ib}(i\frac{ku}{b})}{\mathcal{N}(-\frac{ku}{b})^2 C^{(b)}_{S_2}} \mathsf{A}_{0,3}(p_3,p_4,-\frac{k(u-1)}{b})\mathsf{A}_{0,3}(p_1,p_2,-\frac{ku}{b}).
\end{equation}
Factor of $2$ comes from adding the same contribution from the saddle point $p = - i p_M$. Extra $\mathcal{N}(p) C^{(b)}_{S_2}$ in the denominator comes from the proper normalization of the second three-point function and the $1/t$ factor comes from Gaussian integral. The total prefactor is computed to be
\begin{equation}
 \frac{\rho_b(-\frac{ku}{b}) \rho_{-ib}(i\frac{ku}{b})}{\mathcal{N}(-\frac{ku}{b})^2 C^{(b)}_{S_2}} = -4 \left(\frac{ku}{\pi b}\right)^2,
\end{equation}
so we rewrite this as
\begin{equation}
\Omega_{y\overline{y}} = \frac{2\pi e^{t}}{t} \sum \limits_{k= \frac{1-s}{2}}^{\frac{s-1}{2}} u^2 \Phi_k(u),\,\Phi_k(u) = -\frac{k^2}{\pi^2 b^2} \mathsf{A}_{0,3}(p_3,p_4,\frac{k(u-1)}{b})\mathsf{A}_{0,3}(p_1,p_2,\frac{ku}{b}) \label{phiku}.
\end{equation}
Note that it vanishes at $u=0$ ($x\to 1$) and $u=1$ $(x \to y)$, which is the region we want to restrict the integration to.

\paragraph{Other components of GR 2-form. } From the fact that $\Omega$ is closed we can now find the leading behavior of other components, like $\Omega_{x\overline{x}}$. But first, we want to use that we can restrict to ``spinless'' contributions. If we rewrite everything in polar coordinates
\begin{equation}
\begin{cases}
x = e^{-tu/2} e^{i\theta} \\
y = e^{-t/2} e^{i \psi}
\end{cases},
\end{equation}
all components of $\Omega$ should be $\theta, \psi$ independent (to leading order). Introducing
$$
\Omega_{\pm} = x\overline{y}\, \Omega_{x\overline{y}} \pm y \overline{x} \,\Omega_{y\overline{x}},
$$
the condition that $\Omega$ is closed reads
\begin{equation}
\begin{cases}
\partial_t (te^{-tu} \Omega_{x\overline{x}}) - \partial_u (ue^{-tu} \Omega_{x\overline{x}} + \frac{1}{2}\Omega_+) = 0 \\
\partial_u (e^{-t} \Omega_{y\overline{y}} + \frac{u}{2}\Omega_+) - \frac{1}{2}\partial_t (t \Omega_+) = 0
\end{cases}.
\end{equation}
The second equation gives
\begin{equation}
\Omega_+ = -\frac{4\pi}{t} \sum \limits_k u \Phi_k(u).
\end{equation}
Then from the first one we have, finally,
\begin{equation}
\Omega_{x\overline{x}} |x|^2 = \frac{2\pi}{t} \sum \limits_k \Phi_k(u). \label{omegaxx}
\end{equation}
It remains to pass to logarithmic GR 2-form. 

\paragraph{What changes for logarithmic GR 2-form. } For the logarithmic operator, the leading terms in the OPE of ground ring and intermediate state that we considered differ only by extra HEM $q$-factor and coordinate logarithm $\log (x\overline{x})$. $q$-factor includes the momentum of the resulting state, which at the saddle point reads
\begin{equation}
|p + \frac{k}{b}|= \frac{1}{b}|k|(1-u).
\end{equation}
The coordinate logarithm is $\log x\overline{x} = - t u$. Collecting together \eqref{phiku}, \eqref{omegaxx} and these factors, we finally obtain
\begin{equation}
-\pi \int \limits_0^1 du\,|x|^2\Omega_{x\overline{x}}^{\text{log}}  = \frac{2}{b^3} \int\limits_0^1 du\,u(1-u) \sum \limits_{k=(1-s)/2}^{(s-1)/2} |k|^3 \mathsf{A}_{0,3}(p_3,p_4,\frac{k(u-1)}{b})\mathsf{A}_{0,3}(p_1,p_2,\frac{ku}{b}). \label{HEM-corner-term}
\end{equation}

The last thing to check is that this expression actually reproduces the remainder. Taking the $u$ integral, we obtain
\begin{align*}
&\int \limits_0^1 du\,u\,(1-u) \mathsf{A}_{0,3}(p_1,p_2,(u-1) \frac{k}{b})\mathsf{A}_{0,3}(p_3,p_4,u \frac{k}{b}) = \\
&=4b^2\sum \limits_{m,l=1}^\infty (-1)^{m+l} \frac{\mathcal{N}_{12}(m) \mathcal{N}_{34}(l)}{\sin \pi m b^2 \sin \pi l b^2}  \int \limits_0^1 du\,u\,(1-u) \sin (2\pi m u k)  \sin (2\pi l u k) = \\
& =4b^2\sum \limits_{m,l=1}^\infty (-1)^{m+l} \frac{\mathcal{N}_{12}(m) \mathcal{N}_{34}(l)}{\sin \pi m b^2 \sin \pi l b^2}  \left(\frac{\delta_{m,l}}{12} - \frac{G_1(m,l)}{4\pi^2 k^2} \right) = \\
& = \frac{b^2}{3} \mathsf{C} - \frac{b^2}{k^2} \mathsf{E}_{12|34}
\end{align*}
and the $k$ sum together with the sum over three channels brings precisely
\begin{equation}
\sum \limits_{k=\frac{1-s}{2}}^{\frac{s-1}{2}}|k|^3 \left(b^2 \mathsf{C} - \frac{b^2}{k^2} \mathsf{E}  \right) = b^2 \left(\frac{1}{32} \left(s^2-1\right)^2 \mathsf{C} - \frac{s^2-1}{4} \mathsf{E} \right) = \frac{b^3}{2} \mathsf{R} \label{ksum}.
\end{equation}
The remaining part of the $q$-factor, proportional to $p_{-1,s}$ gives, in fact, an alternative way to compute contributions to the curvature term from $\Delta_{34}$ divisor, discussed previously. Indeed, for this we have the expression of the form
\begin{align}
     &-\frac{2}{b^2}\sum \limits_{k=(1-s)/2}^{(s-1)/2} k^2\int\limits_0^1 du\,u\,\mathsf{A}_{0,3}(p_3,p_4,\frac{k(u-1)}{b})\mathsf{A}_{0,3}(p_1,p_2,\frac{ku}{b}) = \nonumber\\ &=-8 \sum \limits_{k=(1-s)/2}^{(s-1)/2} k^2 \sum \limits_{m,l=1}^\infty (-1)^{m+l} \frac{\mathcal{N}_{12}(m) \mathcal{N}_{34}(l)}{\sin \pi m b^2 \sin \pi l b^2}  \underbrace{\int \limits_0^1 du\,u \sin (2\pi m u k)  \sin (2\pi l u k)}_{\delta_{ml}/4} = \nonumber \\
& =-2 S_2\sum \limits_{m,l=1}^\infty \frac{\mathcal{N}_{1234}(m)}{\sin^2 \pi m b^2}   = -\frac{s(s^2-1)}{6} \mathsf{C} \label{gr-5pt-period}.
\end{align}
We get the answer for GR 2-form period over $\Delta_{34}$ suggested in \eqref{gr2formotherdivisor}. Note that the period over a boundary divisor where ground ring and 1 tachyon bubble off (e.g., $\Delta_{123}$), which is the sum of shifted four-point amplitudes, can formally be written as a similar integral, if for the two-point amplitude we use $\mathsf{A}_{0,2} (p_4,\frac{ku}{b}) = - \frac{1}{2p_4} \delta(p_4-\frac{ku}{b})$.

\subsection{Extension to higher points}
The same considerations allow to explain matrix model expressions for higher-point amplitudes using HEM.
They suggest the following recursion relation for genus zero amplitudes:
\begin{align}
&\mathsf{A}_{0,n+1}(\frac{b+sb^{-1}}{2}, \mathbf{p}) = -\sum \limits_{i=1}^n \sum \limits_{l=\frac{1-s}{2}}^{\frac{s-1}{2}} |p_i-\frac{l}{b}| \mathsf{A}_{0,n}(p_i - \frac{l}{b},\mathbf{p} \setminus p_i)  \nonumber \\
&-4 \sum \limits_{\substack{I_1 \cup I_2 = \lbrace 1,\dots,n \rbrace \\ |I_1|,|I_2| \geq 2}} \sum \limits_{k=1}^{\frac{s-1}{2}} \int \limits_0^{k/b} dq\,q\left(\frac{k}{b}-q\right) \mathsf{A}_{0,|I_1|+1}(q, \mathbf{p}_{I_1}) \mathsf{A}_{0,|I_2|+1}(\frac{k}{b}-q, \mathbf{p}_{I_2}) \nonumber\\
&+p_{-1,s} \Biggl(\sum \limits_{m \neq i,j}^n \sum \limits_{l=\frac{1-s}{2}}^{\frac{s-1}{2}} \mathsf{A}_{0,n}(p_m - \frac{l}{b},\mathbf{p} \setminus p_m) \nonumber \\&+ 4 \sum \limits_{\substack{K \subset \lbrace 1,\dots,n \rbrace \\ i,j \notin K \\
|K| \geq 2}}  \sum \limits_{k=1}^{\frac{s-1}{2}} \int \limits_{0}^{k/b} dq\,q\,\mathsf{A}_{0,|K|+1}(q, \mathbf{p}_{K}) \mathsf{A}_{0,n-|K|+1}(\frac{k}{b}-q, \mathbf{p} \setminus \mathbf{p}_{K}) \Biggr)
 \label{hem-tach-general}.
\end{align}

Some explicit checks of HEM results for $\mathsf{A}_{0,6}$ are performed in Appendix \ref{app-6pf}. There are a few novel subtleties that we comment on below, using this example.
\begin{itemize}
    \item ``Corner'' contributions to the curvature term in $\mathsf{A}_{0,6}$ in HEM calculation come out as integrals of the type \eqref{HEM-corner-term}, where $(1-u)|k|$ or $u|k|$ is replaced by $p_{-1,s}$ and instead of $\mathsf{A}_{0,3}(\dots) \cdot \mathsf{A}_{0,3}(\dots)$ one has $\mathsf{A}_{0,3} (\dots) \cdot \mathsf{A}_{0,4} (\dots)$. The appearing integral is not symmetric anymore with respect to $u \leftrightarrow (1-u)$, unlike the five-point case, so the question is which factor $u$ or $(1-u)$ to erase. 

This is solved individually for each corner, depending on how stationary points $z_1,z_2,z_3$ in our chart are distributed along $(0,3)$ and $(0,4)$ components. The rule is that ground ring ``fuses'' with the component which has $\leq 1$ stationary tachyon (the corresponding component of degenerate curve has vanishing size in the limit) and in the integral we have the corresponding coordinate factor --- $(u-1)$ for $\mathsf{A}_{0,3}$ and $u$ for $\mathsf{A}_{0,4}$. Of 10 total splits, we have $7$ of the first kind and $3$ of the second. 

For general number $n$ of tachyon insertions, based on the considered examples, we can conjecture the following formula for boundary period of ground ring form: for $S \subset \lbrace 1,\dots, n \rbrace$ and $S^C$ the complementary subset
\begin{equation}
\int \limits_{\Delta_{n+1,S} \subset \overline{\mathcal{M}}_{0,n+1}}\Omega_{1,s} = \sum \limits_{l=\frac{1-s}{2}}^{\frac{s-1}{2}} \int \limits_0^{l/b} (-2q\,dq) \mathsf{A}_{0,|S|+1}(\mathbf{p}_S, q)\mathsf{A}_{0,|S^C|+1}(\mathbf{p} \setminus\mathbf{p}_S, q - \frac{l}{b}) \label{gr-period-general}
\end{equation}

    \item Previously the calculation of the curvature term also made use of the result for ``infinity contribution'' $-2\langle O \int U \int U\, TTT \rangle$. However, as before, to compute it one can use a relation in homology: a fiber class $F_{yz}$ of a forgetful map $\overline{\mathcal{M}}_{0,6} \to \overline{\mathcal{M}}_{0,4}$ (four points that are left are $x,z_1,z_2,z_3$) satisfies, for any fixed $i=1,2,3$
\begin{equation}
F_{yz} = \Delta_{xi} + \Delta_{xiy} + \Delta_{xiz} + \Delta_{xiyz}.
\end{equation}
This is analogous to the relation discussed previously in subsection \ref{subs-periods}. Using this, we can see that the total divisor that enters the curvature term will again be dual to $\psi$-class. In \eqref{hem-tach-general}, we rewrote the curvature term using the following representative for $\psi$-class, available in genus zero: for any pair of points $i,j$
$$
\psi_{n+1} = \sum \limits_{K\subset \lbrace 1 \dots n \rbrace,\,i,j \notin K } \delta_{K,n+1}.
$$
    \item Previous two considerations together lead to the conclusion that fixed ground ring period $\mathsf{A}_{0,n|1}$ is not symmetric in integrated and stationary tachyons anymore. For five-point case the difference of periods for different choices of integrated tachyons vanished due to Keel relations, which are more complicated for higher number of points. 
\end{itemize}
\subsection{What happens in $(2,2\mathbf{p}+1)$ minimal string}\label{sec-minimalstringfail}
It is a natural question to try to extend this calculation to rational minimal strings, at least the Lee-Yang series (it was attempted before in \cite{Hamada:1991tz} in a more restricted context, see also \cite{Hamada:1994zv}). In this section we will parametrize conformal dimensions in Liouville by $P$: $\Delta_P = \frac{(b+b^{-1})^2}{4} + P^2$ and in minimal model by $P_M$: $\Delta_{\text{MM}} = - \frac{(b^{-1}-b)^2}{4} + P_M^2$, with $b = \sqrt{\frac{2}{2\mathbf{p}+1}}$; the spectrum in both theories corresponds to real $P, P_M$.

The difference in this case is that instead of the integral over intermediate momenta $P_M$ in the matter sector we have a discrete, finite sum. The finite ``corner'' contribution, which emerged before because of $t^{-1}$ factor from integrals over momenta, now appears in a different way: at fixed $P_M$ we integrate over $P$ and $u$, using saddle point method. We have a quickly decaying exponential factor, given by \eqref{gauss-int} with $b^{-1}$ replaced by $b$. The saddle point is
\begin{equation}
P_\text{sdl} = iP_M,\,u_{\text{sdl}}(m,k)= - \frac{P_M}{bk} = \frac{1}{2k}(\mathbf{p}-m+\frac{1}{2}),
\end{equation}
if the intermediate field in minimal model is $\Phi_{m,1}$. As before, intermediate state goes on shell on the saddle.

The thing that might be confusing is that this calculation gives finite answer only if the $u_{\text{sdl}}$ lies on the integration contour, i.e., between $0$ and $1$. This happens only for intermediate momenta that have
$$
m +2k\geq \mathbf{p}+ \frac{1}{2}.
$$
 For example, for $s=2$ $k \in \lbrace - \frac{1}{2}, \frac{1}{2} \rbrace$ and only one intermediate channel with $m=p$ can contribute. If other parameters of the tachyon correlator are much smaller than $p$, this ``corner'' contribution just does not appear and cannot fix the discrepancy between naive HEM and TR results.

We will explain that this calculation is still not meaningless on the simpler example of 1GR-4T correlator $\langle \mathcal{O}_{k_1,1} \int \mathcal{U}_{k_2,1} \prod \limits_{i=3}^5 \mathcal{T}_{k_i,1} \rangle$\footnote{In this subsection we parametrize tachyons with their matter labels, instead of Liouville momenta, so $U_{k,1} = V_{-k,1} \Phi_{k,1}$.} --- the analysis of five-point tachyon amplitude is more bulky and confusing and we leave it for future work.

For simplicity assume ``generic kinematics'', as in \cite{Artemev:2025pvk}; it means that dimensions of spaces of minimal model conformal blocks/Verlinde numbers are computed as
\begin{equation}
\mathcal{N}^{(0)}_{k_1 k_2 k_3 \dots k_5} = k_1 k_2;\,\mathcal{N}^{(0)}_{k_{1,2} \cdot \cdot \cdot} = k_{1,2};\,\mathcal{N}^{(0)}_{k_{1,2} \,k \,k+l} =
\begin{cases}
  1,\, l =-k_{1,2}+1:2:k_{1,2}-1 \\
  0,\,\text{else}
\end{cases}.
\end{equation}
Additionally assume $k_1 \leq k_2$. The fixed ground ring correlator evaluated via HEM reads (see \cite{Artemev:2025pvk} for the calculation)
\begin{equation}
\mathcal{A}_{0,4|1}(k_2,k_3,k_4,k_5|k_1) = k_1 \alpha_{0,4}(k_2|k_3,k_4,k_5) - \frac{k_1(k_1^2-1)}{3}
\end{equation}
with
$$
\alpha_{0,4}(k_2|k_3,k_4,k_5) =k_2(k_2+\mathbf{p}+\frac{1}{2})- \sum \limits_{i=1}^3 \sum \limits_{l=-k_2+1:2}^{k_2-1} |\mathbf{p}+\frac{1}{2}-k_i-l|.
$$
From the relations in homology we expect that 
\begin{equation}
    \int \limits_{F_2} \Omega_{k_1,1} = \left[\int \limits_{\Delta_{1j}} + \int \limits_{\Delta_{12j}} \right] \Omega_{k_1,1}
\end{equation}
for arbitrary $j \in \lbrace 3,4,5\rbrace$. The first integral is just a sum of shifted four-point functions, which simplifies to:
\begin{equation}
\int \limits_{\Delta_{1j}} \Omega_{k_1,1} = k_1 \alpha_{0,4}(k_2|k_3,k_4,k_5) - \sum \limits_{m=-k_1+1:2}^{k_1-1}\sum \limits_{l=-k_2+1:2}^{k_2-1} (|\mathbf{p}+\frac{1}{2}-k_j-l-m|-|\mathbf{p}+\frac{1}{2}-k_j-l|).
\end{equation}
This means the second period should read
\begin{equation}
    \int \limits_{\Delta_{12j}}\Omega_{k_1,1} = - \frac{k_1(k_1^2-1)}{3} + \sum \limits_{m=-k_1+1:2}^{k_1-1}\sum \limits_{l=-k_2+1:2}^{k_2-1} (|\mathbf{p}+\frac{1}{2}-k_j-l-m|-|\mathbf{p}+\frac{1}{2}-k_j-l|) \label{ms-second-period}.
\end{equation}
The remaining period is also a sum of shifted four-point functions, which simplifies to
\begin{multline}
\int \limits_{\Delta_{12}} \Omega_{k_1,1} = \frac{k_1(k_1^2-1)}{3}+ k_1 k_2 (k_2 +\mathbf{p} + \frac{1}{2}) - \sum \limits_{i=1}^3 \sum \limits_{m=-k_1+1:2}^{k_1-1}\sum \limits_{l=-k_2+1:2}^{k_2-1} |\mathbf{p}+\frac{1}{2}-k_i-l-m|.
\end{multline}
A different quantity was derived from TR in \cite{Artemev:2025pvk} 
\begin{align}
&\mathcal{D}(k_2,k_3,k_4,k_5|k_1) = 2k_1  \alpha_{0,4}(k_2|k_3,k_4,k_5) - \nonumber \\
&-\sum \limits_{i=3,4,5} \sum \limits_{l=-k_2+1:2}^{k_2-1} \sum \limits_{m=-k_1+1:2}^{k_1-1} \left(|\mathbf{p}+\frac{1}{2}-k_i-m-l|-|\mathbf{p}+\frac{1}{2}-k_i-l| \right) \label{diff}
\end{align}
and was proposed to compute ``ground ring correlators''. We suggested earlier that a more precise interpretation for the dictionary of \cite{Artemev:2025pvk} is that it produces the first descendant of GR 2-form. To confirm this, we can use other available periods together with relations in homology and compute $\mathcal{D}$ as (see \eqref{div-dx})
\begin{multline}
\mathcal{D}(k_2,k_3,k_4,k_5|k_1) = \left(\sum \limits_{i=3}^5 \left[\int \limits_{\Delta_{1i}} + \int \limits_{\Delta_{12i}} \right]+ \int \limits_{\Delta_{12}} -2 \int \limits_{F_2}  \right) \Omega_{k_1,1} = \\=\left(\sum \limits_{i=3}^5 \left[\int \limits_{\Delta_{1i}} + \int \limits_{\Delta_{12i}} \right]+ \int \limits_{\Delta_{12}} -2 \left[\int \limits_{\Delta_{1j}} + \int \limits_{\Delta_{12j}} \right]  \right) \Omega_{k_1,1}
\end{multline}
This alternative method reproduces \eqref{diff} precisely. In particular, the $\frac{k_1(k_1^2-1)}{3}$ term cancels.

Note that the second term in \eqref{ms-second-period} is only nonzero for large enough $k_j$; more precisely, if $m+2k>\mathbf{p}+\frac{1}{2}$ for some $m$ that can be obtained as a result of fusion $\Phi_{k_2,1} \Phi_{k_j,1}$ and $k \in \lbrace \frac{1-k_1}{2}, \dots, \frac{k_1-1}{2} \rbrace$. This is precisely the requirement that we found before. In fact, under our assumptions one can check that this term can be rewritten as
\begin{multline}
    \sum \limits_{m=-k_1+1:2}^{k_1-1}\sum \limits_{l=-k_2+1:2}^{k_2-1} (|\mathbf{p}+\frac{1}{2}-k_j-l-m|-|\mathbf{p}+\frac{1}{2}-k_j-l|) = \\ = -2\sum \limits_{m=1}^\mathbf{p} \sum \limits_{k=\frac{1-k_1}{2}}^{\frac{k_1-1}{2}} \mathcal{N}_{k_j k_2 m} \theta(m+2k-\mathbf{p}-\frac{1}{2}) (\mathbf{p}+ \frac{1}{2}-m-2k) = \\ =\sum \limits_{m=1}^\mathbf{p} \sum \limits_{k=\frac{1-k_1}{2}}^{\frac{k_1-1}{2}} 4k\,\mathcal{N}_{k_j k_2 m} \theta(1-u_\text{sdl}(m,k))  (1-u_\text{sdl}(m,k)) ,
\end{multline}
which reminds of a discrete version of the integral over $u$ in \eqref{gr-5pt-period}.

To fix the disagreement between HEM and TR for minimal string, this Polchinski-like calculation is not enough. There must be another source for finite contributions that we have not seen in CLS calculation, in particular, to produce the extra $-\frac{k_1 (k_1^2-1)}{3}$ term in \eqref{ms-second-period} that survives for small $k_j$.

This source indeed exists and is available due to minimal string kinematics. If $k_j$ is small, discrete terms appear in the Liouville OPE of $V_{-k_2,1} V_{-k_j,1}$. They have momenta
$$
P_{\text{discrete}} = - \frac{ib}{2}\left(\mathbf{p}+\frac{1}{2}-(k_2 + k_j +2n+1) \right),\,n \in \mathbb{Z}_{\geq 0}
$$
for $n$ such that expression in brackets is $>0$ (one can also shift by $b^{-1}$, but for large enough $p$ this possibility can be ignored). In minimal model OPE the fields $\Phi_{k,1},\,k \in \lbrace |k_j-k_2|+1:2:k_j+k_2-1 \rbrace$ appear. These can not combine into on-shell tachyon with the listed discrete terms. However, if additionally the OPE with degenerate fields in both sectors is performed, it is possible to go on shell; e.g., if we additionally introduce $V_{2,1} \Phi_{2,1}$, one gets the channel $\Phi_{k_j+k_2,1}$ in the matter sector, and after OPE with discrete term we can obtain $P = -\frac{ib}{2}\left(\mathbf{p}+\frac{1}{2}-k_2-k_j \right)$ in Liouville. The count of terms where Liouville and matter operators combine into on-shell tachyon for small $k_1$ gives $\frac{k_1(k_1^2-1)}{6}$: $1$ for $k_1=2$, $4$ for $k_1 =3$ and so on. Thus if these discrete terms are taken with equal coefficients, they might indeed produce the first term in \eqref{ms-second-period}.

\section{Intersection-theoretic formula for ground ring insertions} \label{sec: inttheorygr}
It is natural to look for expressions for string ``amplitudes'' with ground ring insertions (which are non-top degree forms on corresponding moduli spaces) in terms of standard tautological classes. In this section we will propose how to construct such an expression, starting from the tautological class that describes tachyon operators, proposed in \cite{Collier:2024lys}. Perhaps it would allow to find a dual matrix model/TR description for them, but for now the exact meaning of this in the dual setup is not clear.  
\subsection{Motivation: the unit operator and pullback}
For convenience we repeat the intersection-theoretic formula for the tachyon class
\begin{multline}
\mathsf{A}_{g,n} = \int \limits_{\overline{\mathcal{M}}_{g,n}}\sum \limits_{\Gamma} \frac{1}{|\text{Aut\,}\Gamma|} \iota_{\Gamma *} \Theta_\Gamma,\, \\ \Theta_\Gamma = \prod \limits_e \left(\int' -2q_e dq_e\,e^{-q_e^2 (\psi_{e\circ} + \psi_{e\bullet})} \right) \prod \limits_{l} e^{-p_l^2 \psi_l} \prod \limits_v \mathcal{N}_v\, e^{\sum \limits_{i=1}^\infty t_i \kappa_i^v}.
\end{multline}
To motivate the prescription that will follow, we first consider the simplest case of unit operator $\mathcal{O}_{1,1}$, inserted in the tachyon $n$-point function. Then the corresponding class on $\overline{\mathcal{M}}_{g,n+1}$ is just a pullback of the tachyon class under the forgetful map.

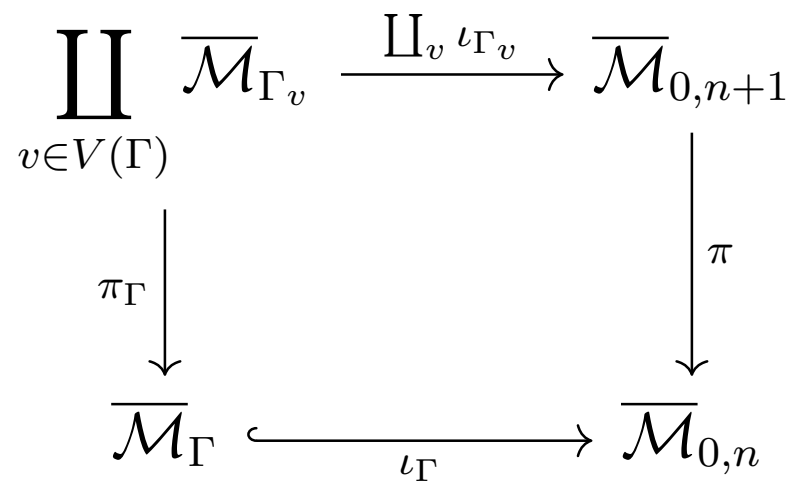
\begin{figure}
    \centering
   \begin{tikzcd}[column sep=large, row sep=large]
\displaystyle\coprod_{v\in V(\Gamma)} \overline{\mathcal M}_{\Gamma_v}
  \arrow[r, "\coprod_v \iota_{\Gamma_v}"] \arrow[d, "\pi_\Gamma"']
& \overline{\mathcal M}_{0,n+1} \arrow[d, "\pi"] \\
\overline{\mathcal M}_{\Gamma} \arrow[r, hook, "\iota_\Gamma"']
& \overline{\mathcal M}_{0,n}
\end{tikzcd}
        \caption{Maps between moduli spaces}
    \label{maps}
\end{figure}

 Due to commutative diagram on Figure \ref{maps} we can instead consider pullback of each graph contribution to $\overline{\mathcal{M}}_{\Gamma_v}$ --- moduli space parametrized by a stable graph with $n+1$ legs and $n+1$-st leg attached to some vertex $v$ of $\Gamma$. 
Pullback under the forgetful map reads
\begin{align*}
\pi^* \kappa_k = \kappa_k -\psi_{n+1}^k,\\
\pi^* \psi_i^d = \psi_i^d + (- \delta_{i,n+1})^{d} .
\end{align*}
So the factors in $\Theta_{\Gamma}$ are pulled back to $\coprod \overline{\mathcal{M}}_{\Gamma_v}$ as
\begin{equation}
e^{\sum t_k \kappa_k} \to e^{\sum t_k \kappa_k}  \left(e^{-\sum t_k\psi_{n+1}^k} = \frac{1}{\psi_{n+1}} \left(e^{-\frac{\hat{(Q-2b)}^2}{4} \psi_{n+1}} - e^{-\frac{Q^2}{4}\psi_{n+1}} \right) \right),
\end{equation}
\begin{equation}
e^{-q^2 \psi_i} \to e^{-q^2 \psi_i} + e^{q^2 \delta_{i,n+1}}-1
.\end{equation}
The correction in the second formula we can further express as 
\begin{equation}
e^{q^2 \delta_{i,n+1}}-1 = \rho_*\left(\frac{e^{-q^2 \psi_{\bullet}}-1}{-\psi_{\bullet}} \right)
\end{equation}
where $\rho$ is the inclusion map for moduli space associated to a refined graph, which has an edge separating vertex $v$ and a new vertex where legs $i$ and $n+1$ enter; $\psi_\circ$ and $\psi_\bullet$ are associated to this edge. Note that $\psi_{\circ}=0$, since the other half edge is associated to a $(g,n) =(0,3)$ component. All this is valid when $i$ is an external leg or an internal half-edge. For internal edges by $\delta_{i,n+1}$ we mean the divisor where $(n+1)$-st point meets the associated node.

\begin{figure}[ht]
\centering
\begin{subfigure}[b]{0.25\linewidth}
\centering
\begin{tikzpicture}[tg]
  \foreach \x/\nl in {0/0, 3.1/1} {
    \begin{scope}[shift={(\x,0)}]
      \node[vert] (v) at (0,0) {};
      \node[sm, below left=0pt] at (v) {$v$};
      \draw[edge] (v) -- (-0.9,0.5); \draw[edge] (v) -- (0.85,0.75); \draw[edge] (v) -- (0.85,-0.4);
      \ifnum\nl=1 \draw[legn] (v) -- (0,0.9) node[sm, above, red!80!black] {$n{+}1$}; \fi
      \blob{(-0.9,0.5)} \blob{(0.85,0.75)} \blob{(0.85,-0.4)}
    \end{scope}}
  \node[sm] (G)  at (0,-1.3)   {$\Gamma$};
  \node[sm] (Gv) at (3.1,-1.3) {$\Gamma_v=\hat\Gamma$};
  \draw[mapto] (Gv) -- node[sm, above] {$\pi$} (G);
\end{tikzpicture}
\caption{Type 1}
\end{subfigure}\hfill
\begin{subfigure}[b]{0.37\linewidth}
\centering
\begin{tikzpicture}[tg]
  \node[vert] (v) at (0,0) {};
  \draw[edge] (v) -- (-0.85,0.45); \draw[edge] (v) -- (0.9,0);
  \draw[legi] (v) -- (0,0.9) node[sm, above, blue!70!black] {$i$};
  \blob{(-0.85,0.45)} \blob{(0.9,0)}
  \begin{scope}[shift={(2.9,0)}]
    \node[vert] (v) at (0,0) {};
    \draw[edge] (v) -- (-0.85,0.45); \draw[edge] (v) -- (0.9,0);
    \draw[legi] (v) -- (0,0.9) node[sm, above, blue!70!black] {$i$};
    \draw[legn] (v) -- (0.55,0.85) node[sm, above right=-1pt, red!80!black] {$n{+}1$};
    \blob{(-0.85,0.45)} \blob{(0.9,0)}
  \end{scope}
  \begin{scope}[shift={(5.8,0)}]
    \node[vert] (v) at (0,0) {}; \node[vert] (u) at (0.05,0.6) {};
    \draw[edge] (v) -- (-0.85,0.45); \draw[edge] (v) -- (0.9,0);
    \draw[newedge] (v) -- (u);
    \draw[legi] (u) -- (-0.2,1.15) node[sm, above, blue!70!black] {$i$};
    \draw[legn] (u) -- (0.55,1.0) node[sm, above right=-1pt, red!80!black] {$n{+}1$};
    \blob{(-0.85,0.45)} \blob{(0.9,0)}
  \end{scope}
  \node[sm] (G)  at (0,-1.3)   {$\Gamma$};
  \node[sm] (Gv) at (2.9,-1.3) {$\Gamma_v$};
  \node[sm] (Gh) at (5.8,-1.3) {$\hat\Gamma$};
  \draw[mapto]  (Gv) -- node[sm, above] {$\pi$} (G);
  \draw[hookto] (Gh) -- node[sm, above] {$\rho$} (Gv);
\end{tikzpicture}
\caption{Type 2}
\end{subfigure}\hfill
\begin{subfigure}[b]{0.37\linewidth}
\centering
\begin{tikzpicture}[tg]
  \node[vert] (v) at (-0.28,0) {};
  \draw[edge] (v) -- (0.75,0);
  \blob{(-0.6,0)} \blob{(0.75,0)}
  \node[vert] at (v) {};
  \node[sm, below=2pt] at (-0.1,-0.05) {$v$};
  \begin{scope}[shift={(2.9,0)}]
    \node[vert] (v) at (-0.28,0) {};
    \draw[edge] (v) -- (0.75,0);
    \blob{(-0.6,0)} \blob{(0.75,0)}
    \draw[legn] (v) -- (-0.28,0.85) node[sm, above, red!80!black] {$n{+}1$};
    \node[vert] at (v) {};
  \end{scope}
  \begin{scope}[shift={(5.8,0)}]
    \node[vert] (u) at (0.1,0) {};
    \draw[newedge] (-0.75,0) -- (u);
    \draw[edge] (u) -- (0.95,0);
    \blob{(-0.75,0)} \blob{(0.95,0)}
    \draw[legn] (u) -- (0.1,0.85) node[sm, above, red!80!black] {$n{+}1$};
    \node[vert] at (u) {};
  \end{scope}
  \node[sm] (G)  at (0,-1.3)   {$\Gamma$};
  \node[sm] (Gv) at (2.9,-1.3) {$\Gamma_v$};
  \node[sm] (Gh) at (5.8,-1.3) {$\hat\Gamma$};
  \draw[mapto]  (Gv) -- node[sm, above] {$\pi$} (G);
  \draw[hookto] (Gh) -- node[sm, above] {$\rho$} (Gv);
\end{tikzpicture}
\caption{Type 3}
\end{subfigure}
\caption{The three positions of the leg $n{+}1$.}
\label{figure-types}
\end{figure}
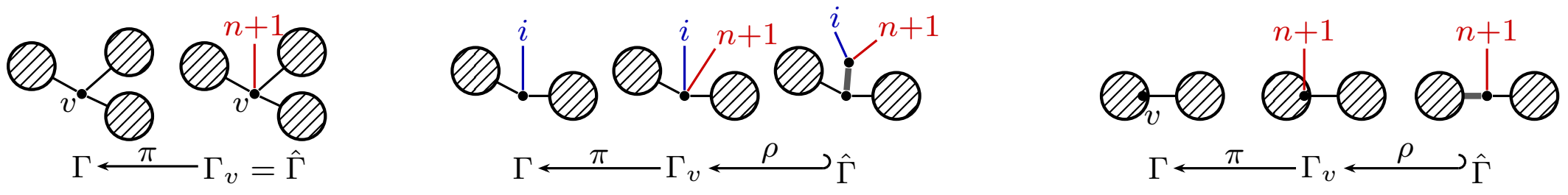

A replacement of each factor, in fact, ``linearizes'' --- $\psi_{n+1}$ does not intersect with $\delta_{i,n+1}$ for any $i$ and also $\delta_{i,n+1} \delta_{j,n+1} = 0$ for $i \neq j$, so we can just sum over additions to each vertex and half-edge. For vertex correction we multiply by
$$
\frac{1}{\psi_{n+1}} \left(e^{-\frac{(Q-2b)^2}{4} \psi_{n+1}} - e^{-\frac{Q^2}{4}\psi_{n+1}} \right)
$$
and for half-edge we replace it by
$$
e^{q^2 \delta_{i,n+1}}-1.
$$
We can thus express the pullback as a sum over graphs $\hat{\Gamma}$ of type $(g,n+1)$, where a new $n+1$ leg can appear in three ways:
\begin{enumerate}
    \item attached to vertex $v$ of graph $\Gamma$ ($\kappa$-class corrections);
    \item fusing with leg $i$ into three-valent vertex ($\psi_i$-class corrections);
    \item breaking the internal edge into two parts (correction to one of the exponents in the edge factors).
\end{enumerate}
The three options are depicted in Figure \ref{figure-types}.

As the most basic check, let us compute the integral over the divisor $\Delta_{i,n+1}$. Obviously only graphs $\Gamma_v$ where $n+1$ and $i$ enter the same vertex $v$ contribute. For each such graph $\Gamma_v$
\begin{equation}
\int \limits_{\overline{\mathcal{M}}_{0,n+1}} \delta_{i,n+1} \,\iota_{\Gamma_v *} \Theta_{0,n|1} = \int \limits_{\overline{\mathcal{M}}_{\Gamma_v}} \Theta_{0,n|1} \,\iota_{\Gamma_v}^* \delta_{i,n+1}
\end{equation}
because of projection formula. The graphs $\Gamma_v$ are such that vertex $v$ necessarily has $\geq 4$ entering legs/edges (this is because they were stable before $(n+1)$-st leg was added). Then one can write
\begin{equation}
\iota_{\Gamma_v}^* \delta_{i,n+1} = \rho_*1
\end{equation}
with $\rho: \overline{\mathcal{M}}_{\hat{\Gamma}} \to \overline{\mathcal{M}}_{\Gamma_v}$ an inclusion map defined above. Using the projection formula again,
\begin{equation}
\int \limits_{\overline{\mathcal{M}}_{\Gamma_v}} \Theta_{0,n|1} \,\iota_{\Gamma_v}^* \delta_{i,n+1} = \int \limits_{\overline{\mathcal{M}}_{\hat{\Gamma}}} 1\cdot  \rho^*\Theta_{0,n|1}.
\end{equation}
Pullback does not touch kappa-classes, nullifies $\psi_i$ and $\psi_{n+1}$ and gives $-\psi_\bullet$ for $\delta_{i,n+1}$. So, it acts trivially almost on everything except
\begin{equation}
\rho^*\left(    e^{-p_i^2 \psi_i} + e^{p_i^2 \delta_{i,n+1}}-1 \right) = e^{-p_i^2 \psi_\bullet}.
\end{equation}
If we present our class as a sum over graphs $\hat{\Gamma}$, where only single element is changed at a time, what contributes are type $1$ graphs with $i$, $n+1$ entering the same vertex and type $2$ graphs where $(n+1)$-st leg fuses with $i$. Type $1$ graphs contribute $(+1)$ for each choice of vertex $v$, while for type $2$ we use
\begin{equation}
\rho^* \rho_*\left(\frac{e^{-q^2 \psi_{\bullet}}-1}{-\psi_{\bullet}} \right) = e^{-q^2 \psi_\bullet} - 1.
\end{equation}
Note that $\overline{\mathcal{M}}_{\hat{\Gamma}_v} $ for graphs of type $2$ is isomorphic to $\overline{\mathcal{M}}_{\Gamma}$ (evident from the picture), with $\psi_\bullet$ replacing $\psi_i$. The total sum gives the same as the tachyon $n$-point correlator we started with, as expected.

\subsection{Dictionary} \label{sec: gr-inttheory-dictionary}

For the insertion of GR operator $\mathcal{O}_{1,s}$ with $s \neq 1$ in the tachyon amplitudes we also want to write down some operation that presents the result as sum over $\hat{\Gamma}$, which in some sense is a ``modified pullback''. The prescription we propose is as follows: take the contribution to tachyon class associated to a graph $\Gamma$ of type $(0,n)$ and for each refinement $\hat{\Gamma}$, described in the previous section, modify it as written below; the result is interpreted as a class on $\overline{\mathcal{M}}_{\hat{\Gamma}}$.
\begin{enumerate}
    \item For graphs of type $1$, replace the vertex factor associated with $v$
    $$
    e^{\sum t_k \kappa_k} \to e^{\sum t_k \kappa_k}  \left(\frac{1}{\psi_{n+1}} \left(e^{-p_{-1,s}^2 \psi_{n+1}} - e^{-p_{1,s}^2\psi_{n+1}} \right) \right).
    $$
    \item For graphs of type $2$, replace the leg factor
    $$
    e^{-p_i^2 \psi_i} \to \sum \limits_{l = \frac{1-s}{2}}^{\frac{s-1}{2}} \frac{1-e^{-(p_i-l/b)^2 \psi_\bullet}}{\psi_\bullet}.
    $$
    \item For graphs of type $3$, replace the edge factors  with the integrated momentum
    $$
    e^{-q^2 (\psi_\circ + \psi_\bullet)} \to \sum \limits_{l = \frac{1-s}{2}}^{\frac{s-1}{2}} \left(\frac{1-e^{-(q-l/b)^2 \psi_\circ}}{\psi_\circ} e^{-q^2 \psi_\bullet} +  \frac{1-e^{-(q-l/b)^2 \psi_\bullet}}{\psi_\bullet} e^{-q^2 \psi_\circ} \right) .
    $$
\end{enumerate}
Note that type $1$ dressing can be presented similarly to type $2$ as the action on the difference
\begin{equation}
\left(e^{-\frac{Q^2\psi}{4}}- e^{-\frac{\hat{Q}^2\psi}{4}}\right) \to \sum \limits_{l = \frac{1-s}{2}}^{\frac{s-1}{2}}  \left(\frac{1-e^{-(\frac{Q}{2}-l/b)^2 \psi_{n+1}}}{\psi_{n+1}} -\frac{1-e^{-(\frac{\hat{Q}}{2}-l/b)^2 \psi_{n+1}}}{\psi_{n+1}}\right).
\end{equation}
Indeed, in this sum most of the terms cancel and only two exponentials are left. 

After that modification, each class is pushed forward to $\overline{\mathcal{M}}_{0,n+1}$ by the inclusion map and sum over $\hat{\Gamma}$. Then a term of appropriate degree in cohomology is taken.

\subsection{Computation of boundary periods } To check the proposal, we would need to confirm that it produces correct boundary periods, or intersection numbers with delta-classes $\delta_S$, $S \subset \lbrace 1,\dots,n+1 \rbrace$. They are given either by a sum of shifted tachyon $n$-point functions or integrals of the type \eqref{gr-period-general}.

We do not have a general proof, but for each boundary period the check is straightforward. The calculation strategy is as follows: consider some fixed graph $\hat{\Gamma}$ in the sum and denote the class on $\overline{\mathcal{M}}_{\hat{\Gamma}}$, obtained with the prescription above, as $\Theta$. It will only contribute to the intersection number with some $\delta_S$ if there is a common refinement for $\hat{\Gamma}$ and the one-edge graph $\Gamma_{\Delta_S}$, representing the boundary divisor $\Delta_S$. For such graphs we first use the projection formula and rewrite the contribution to intersection number as follows
\begin{equation}
\int \limits_{\overline{\mathcal{M}}_{0,n+1}} \delta_S\cdot \iota_{\hat{\Gamma}*} \Theta = \int \limits_{\overline{\mathcal{M}}_{\hat{\Gamma}}} \iota_{\hat{\Gamma}}^*\delta_S \cdot  \Theta.
\end{equation}
Then there are two options. If $\hat{\Gamma}$ already contains an edge $E_S$ separating $S$ from other legs (i.e., it is itself a common refinement of $\hat{\Gamma}$ and $\Gamma_{\Delta_S}$), then we have an ``excess intersection''. Pullback of boundary class is $\iota_{\hat{\Gamma}}^*\delta_S = - \psi_\circ - \psi_\bullet$, where the two psi-classes are associated to two half-edges of $E_S$. In the other case an edge needs to be added to pass to common refinement $\hat{\Gamma}_S$. Denote $\rho: \overline{\mathcal{M}}_{\hat{\Gamma}_S}\xhookrightarrow{}\overline{\mathcal{M}}_{\hat{\Gamma}}$, then $\iota_{\hat{\Gamma}}^*\delta_S = \rho_*1$ and we can use the projection formula again:
\begin{equation}
\int \limits_{\overline{\mathcal{M}}_{\hat{\Gamma}}} \iota_{\hat{\Gamma}}^*\delta_S \cdot  \Theta = \int \limits_{\overline{\mathcal{M}}_{\hat{\Gamma}_S}} 1 \cdot  \rho^*\Theta.
\end{equation}
The pullback is then simple to compute. 

In particular, for divisors of the type $\Delta_{i,n+1}$ we need to consider either type $2$ graphs with $(n+1)$-st leg attached to $i$, or type $1$ graphs, where both legs $n+1,i$ are attached to the same vertex $v$. The calculation is fully analogous to the case of unit operator (previous section) and we reproduce a sum of shifted tachyon $n$-point functions as the corresponding period. Divisors when $(n+1)$-st point is together with more than one points of tachyon insertion are more complicated: more graphs contribute, including graphs of type $3$ if $n$ is large enough. For this situation the proposal in the previous section is tested on the simplest example of 1GR-5T amplitude in Appendix \ref{app-boundperiod}. Although this case requires only zero-degree terms in the replaced edge factor, it is still a nontrivial check that is passed by our prescription.
\section{Discussion} \label{sec: discussion}
\paragraph{More precise relation to TR. } The most ambitious goal would be, of course, to establish TR relations on amplitudes directly from the worldsheet (at least for some set of momenta which would allow to extend them to generic values by analytic continuation). This has not been achieved yet.
The HEM recursion that we discussed does not coincide verbatim with the one that follows from TR when we choose one operator to be degenerate. It presents the amplitude as a sum of a similar form and some ingredients of degenerate TR recursion kernels, described in appendix \ref{app-degen-rec}, are the same as in HEM formulas, but others are not. Most significant difference is that the curvature term is distributed differently --- HEM does not treat boundary divisors in a symmetric way\footnote{We could, of course, rewrite the curvature term using a more symmetric representative for $\psi$-class, but this would lead to coefficients in the formula that depend explicitly on $n$, the number of insertions.}. Thus the questions of what is the precise relation between two recursion formulas and whether HEM results can be leveraged to understand more about amplitudes with generic momenta are open. 

\paragraph{(Rational) minimal string generalization. } As mentioned in the text, rational minimal string (at least the $(2,2\mathbf{p}+1)$ case, where the hypothetical dual is most understood; see \cite{Belavin:2008kv, Artemev:2025pvk}) is where HEM method should have the most power, since all amplitudes in this theory include degenerate tachyons. We believe that all possible sources that contribute to the amplitude have now been identified and HEM should reproduce the answers from the dual approach in this theory as well. However, as of now, each specific amplitude requires accurate bookkeeping and it is difficult to write down a completely general answer (even under ``genericity assumptions''). We leave more accurate analysis of tachyon correlators in minimal string for future work.

\paragraph{Relation to ghost dilaton theorem.} The paper of Polchinski, on which the calculation of ``corner terms'' in section \ref{sec-codim-2} is based, presents ``cohomological HEM'' relations in an interesting way, slightly different from ours. Adapted to our notations and generalized, they can be written as
\begin{equation}
B_{r,s} T_{r,-s} =  2p_{r,s} \mathfrak{C}\,O_{r,s}-Q_{\text{BRST}} \left(\mathfrak{c} D_{r,s} \overline{H}_{r,s} V'_{r,s} \Phi_{r,s} \right) + (Q_{\text{BRST}}\text{-exact}) \label{PolchinskiHEM}
\end{equation}
with 
$$
\mathfrak{C} = \sum \limits_{n=1}^\infty n \mathfrak{c}_n \mathfrak{c}_{-n}.
$$
Consider the first term, proportional to $p_{r,s}$. This operator is of ghost number $2$, but only in a holomorphic sector, and formally BRST-exact, because
$$
2\mathfrak{C} = \lbrace Q_{\text{BRST}}, \mathfrak{c}_0 \rbrace.
$$
 However, it is still nontrivial in semirelative cohomology \cite{Polchinski:1998rq} that one usually considers in string theory: its ``BRST ancestor'' $\mathfrak{c}_0 O_{r,s}$ is not annihilated by $(\mathfrak{b}_0- \overline{\mathfrak{b}}_0)$. So this operator nontrivially affects the amplitudes (the remainder denoted as ``$Q$-exact'' in \eqref{PolchinskiHEM} does not have this problem). In the simplest case $(r,s)=(1,1)$ operator $\mathfrak{C} O_{1,1}$ reduces to $\mathfrak{c} \,\partial^2 \mathfrak{c}$ and is known as the ``ghost dilaton''. A subtle computation shows that its insertion in the amplitude reduces to some specific boundary terms and in the end multiplies it by the Euler characteristic \cite{Distler:1990ea}. One can think of associated integrated operator as the Ricci curvature on the string worldsheet, for some (essentially arbitrary) choice of the metric \cite{Polchinski:1988jq}. The interpretation is that this operator is associated to variations of the string coupling (for more precise statement in the context of string field theory, known as the ``dilaton theorem'', see \cite{Bergman:1994qq}). Equations for $(r,s) \neq (1,1)$ produce a natural generalization of ghost dilaton, built from degenerate fields on higher levels; we will call them ``higher ghost dilaton'' operators. 

Two terms in equation \eqref{PolchinskiHEM} can be matched to two contributions in the full HEM answers: the curvature term and everything else (terms, containing ``absolute values'' of intermediate momenta). It is natural to conjecture that ``higher ghost dilaton'' part contributes only to the former. It would be interesting to check this by an explicit calculation, which would reformulate the calculation of curvature term as a higher analogue of the dilaton theorem.

\paragraph{TR interpretation for the ground ring. } With the intersection theory formula of section \ref{sec: inttheorygr}, we described a general answer for ground ring class on the moduli space, but we have not yet found a nice interpretation for this class in TR terms\footnote{The author thanks Sergey Shadrin and Maxim Kazarian for discussions on this point. }. This would be an interesting extension of dictionary of the duality. One hint in this direction is the formula for boundary periods \eqref{gr-period-general}, which suggests that in terms of TR $n$-differentials each such period is computed by a residue transformation of their bilinear combination, ``glued'' in a common point with some weight. 
\section*{Acknowledgements}
The author thanks Alexey Litvinov, Dmitry Khromov, Igor Chaban and Andrei Marshakov for numerous discussions over the years and interest in the topic. He is especially grateful to Lorenz Eberhardt for helpful discussions and suggestions that finally allowed the author to make some progress on this project. The author acknowledges the use of Claude Opus 5.5 for cross-checking calculations, producing figures for the paper and literature exploration. The author takes full responsibility for all results presented in this work. The author was supported by the European Research Council (ERC) under the European Union’s Horizon 2020 research and innovation programme (grant agreement No 101115511).

\appendix

\section{Specialization of recursion for degenerate tachyon} \label{app-degen-rec}
We start with (3.43) of \cite{Collier:2024lys}. It is more convenient since we can simply substitute degenerate momentum $p_1 = \frac{b + sb^{-1}}{2}$ without any problems, unlike in the integral form of the recursion. The recursion has the form
\begin{multline}
p_1 \mathsf{A}_{0,n} = \frac{\pi}{2} \underset{u=0}{\text{Res\,}} \Bigl( \frac{\sin \frac{2\pi u}{b} (s+b^2)}{\sin 2\pi b u \sin 2\pi b^{-1}u} \Bigl[-2\sum \limits_{j=2}^n \int2q\,dq\,\cos 4\pi u q\cos 4\pi u p_j \times \\ \times \mathsf{A}_{0,3} (p_1,p_j,q) \mathsf{A}_{0,n-1}(q,p \backslash p_j) +\int 2qdq\,2q'dq' \mathsf{A}_{0,3}(q,q',p_1) \cos 4\pi u q\cos 4\pi u q' \\\sum \limits_{I+J= \lbrace 1 \dots n \rbrace} \mathsf{A}_{0,1+|I|}(q,p_I)\mathsf{A}_{0,1+|J|}(q',p_J)  \Bigr] \Bigr).
\end{multline}
We want to make use of the following structure of the amplitudes: as a function of one of the momenta (say, $q$)
\begin{equation}
\mathsf{A}_{0,m}(q,\dots) = \sum \limits_{k \geq 0} q^{2k}  A^m_k   = \sum \limits_{k \geq 0, n \geq 1} q^{2k} \sin (2\pi n b q) a_{k,n}^m .
\end{equation}
We will omit upper index $m$ when it is clear from the context. Moreover, three-point amplitude with degenerate $p_1$ specializes as follows (for odd $s$)
\begin{equation}
\mathsf{A}_{0,3}(p_{1,s},p,q) = b\sum \limits_{m \in \mathbb{Z}} \sin (2\pi m b p) \sin (2\pi m b q) .
\end{equation}
Consider the first term in the recursion. Expanding $\mathsf{A}_{0,n-1}$ as a function of $q$, $\mathsf{A}_{0,3}$ as a sum over $m$ and integrating over $q$ term by term, we obtain under the residue
\begin{multline}
-4b \cos (4\pi u p_j) \sum \limits_{m \in \mathbb{Z}} \sin 2\pi m b p_j \sum \limits_{k \geq 0, n\geq 1} \sum \frac{\Gamma(2k+2)}{4(2\pi i b)^{2k+2}} \times \\ \times \sum \limits_{\pm} \left( \frac{1}{((m-n) \pm 2u/b)^{2k+2}} - \frac{1}{((m+n) \pm 2u/b)^{2k+2}} \right).
\end{multline}
Sum over $m$ can be performed now
\begin{multline}
 \sum \limits_{m \in \mathbb{Z}} \sin 2\pi m b p_j\left( \frac{1}{((m-n) \pm 2u/b)^{2k+2}} - \frac{1}{((m+n) \pm 2u/b)^{2k+2}} \right) = \\
 =\sum \limits_{m' \in \mathbb{Z}} \frac{1}{(m' \pm 2u/b)^{2k+2}} \left[\sin 2\pi (m'+n) b p_j - \sin 2\pi (m'-n) b p_j \right] =\\ = 4 \sin 2\pi n bp_j \sum \limits_{m' \in \mathbb{Z}} \frac{\cos 2\pi m' bp_j}{(m' + 2u/b)^{2k+2}} .
\end{multline}
$\pm$ terms give the same, which produces additional factor of $2$ in the end. Denoting $2u/b=t$, the sum is periodic in $x=bp_j$ with period $1$ and for $x \in (0,1)$ can be computed as
\begin{multline}
\sum \limits_{m' \in \mathbb{Z}} \frac{\cos 2\pi m' x}{(m' + t)^{2k+2}} = \frac{1}{\Gamma(2k+2)} (-\partial_t)^{2k+1} \sum \limits_{m' \in \mathbb{Z}} \frac{\cos 2\pi m' x}{(m' + t)} = \\=\frac{-1}{\Gamma(2k+2)} \partial_t^{2k+1} \frac{\pi \cos  \pi t(1-2x)}{\sin \pi t},\,x=bp_j\text{ mod }1.
\end{multline}
Returning to the start, we have
\begin{multline}
-\cos (2\pi t bp_j) b \sum \limits_{k \geq 0, n \geq 1}  \sin 2\pi n x\,  \,\underbrace{\frac{4 (-1)^k}{(2\pi b)^{2k+2}} \partial_t^{2k+1} \frac{\pi \cos  \pi t(1-2x)}{\sin \pi t}}_{\Lambda_k(x,t)} a_{k,n}  = \\ = - b \cos (2\pi t bp_j) \sum \limits_{k \geq 0} A_k(p_j) \Lambda_k(x,t).
\end{multline}
where $A_k$ is the Fourier sum for the amplitude $\mathsf{A}_{0,n-1}$ at fixed polynomial order $k$. Collecting all the coefficients, for the first term in the recursion we can obtain
\begin{equation}
\sum \limits_{k \geq 0}\underbrace{\left(- \frac{\pi b^2}{4}\right) \underset{t=0}{\text{Res\,}} \frac{\Lambda_k(x,t) \sin \pi t(s+b^2)\cos 2\pi t bp_j}{\sin \pi t \sin \pi t b^2}}_{J_k(bp_j)} A_k(p_j).
\end{equation}
Similarly one can transform the separating term; it assumes the form
\begin{equation}
\int \limits_0^1 du\,\sum \limits_{k_1,k_2 \geq 0} A^I_{k_1}(u/b) A^J_{k_2}(u/b) \cdot \underbrace{\frac{\pi b^2}{16} \underset{t=0}{\text{Res\,}}\Lambda_{k_1}(u,t) \Lambda_{k_2}(u,t) \frac{\sin \pi t(s+b^2)}{\sin \pi t \sin \pi t b^2}}_{H_{k_1 k_2}}.
\end{equation}
For illustration we compute explicitly $J_k$ with $k \leq 2$ and $H_{k_1 k_2}$ with $(k_1,k_2) = (0,0)$ or $(0,1)$. Computing
\begin{equation}
\frac{J_k}{p_{1,s}} = - \sum \limits_{l = \frac{1-s}{2}}^{\frac{s-1}{2}}|p_j-\frac{l}{b}|^{2k+1} + p_{-1,s} 
\begin{cases}
    \frac{s}{3},\,k=0 \\
    \sum \limits_{l=\frac{1-s}{2}}^{\frac{s-1}{2}} (p_j-\frac{l}{b})^2-\frac{2S_2}{5 b^2} - \frac{2s t_1}{15},\, k =1 \\
    \sum \limits_{l=\frac{1-s}{2}}^{\frac{s-1}{2}} (p_j-\frac{l}{b})^4 +\frac{4S_2t_1}{7b^2} + \frac{2S_4}{7b^4}+ \\+\frac{s(16t_1^2-1)}{63}-\frac{4t_1 s p_j^2}{3}+\frac{2s}{3} p_j^4,\,k =2
\end{cases}. \label{Jk}
\end{equation}
Here $S_k = \sum \limits_{l=\frac{1-s}{2}}^{\frac{s-1}{2}}|l|^k
$, $t_1 = \frac{b^2 + b^{-2}}{4}$. These formulas can be checked when $|bp_j|<1$ and also extend outside of this region. Substituted into the recursion, the first term gives
\begin{equation}
-\sum \limits_l |p_j - \frac{l}{b}|\sum \limits_{k \geq 0} (p_j-l/b)^{2k} A_k(p_j) = -\sum \limits_l |p_j - \frac{l}{b}| \mathsf{A}_{0,n-1}(p_j-\frac{l}{b}) ,
\end{equation}
which is the expected HEM contribution. The other sum we wrote in the remainder of \eqref{Jk} gives a ``codimension-1'' contribution to the curvature term --- sum of shifted $(n-1)$-point amplitudes with  $p_{-1,s}$ as a coefficient.

The coefficients for separating piece evaluate to
\begin{multline}
H_{00} = B_2(u)\left(-\frac{S_1}{b^3} + \frac{s p_{-1,s}}{3b^2} \right) + \left(\frac{S_3}{6b^3} +p_{-1,s} \left(s\frac{t_1}{45} - \frac{S_2}{10b^2} \right)  \right), \\
H_{01} = B_4(u) \left(-\frac{S_1}{b^5}+ \frac{s p_{-1,s}}{3b^4} \right) + B_2(u) \left(-\frac{S_3}{2b^5} + p_{-1,s} \left(-\frac{s t_1}{15b^2} + \frac{3S_2}{10b^4} \right) \right) + \\+ \left(\frac{S_5}{20b^5} + p_{-1,s} \left(-\frac{S_4}{28b^4} +\right. \right. 
+\frac{S_2}{35} \frac{t_1}{b^2} - s \frac{16t_1^2-1}{1260} \Bigr)\Bigr).
\end{multline}
where $B_k(u)$ are Bernoulli polynomials.

\section{Degenerate tachyon 6-point function} \label{app-6pf}
The formula for degenerate $6$-point amplitude can be produced by specialization of recursion in \cite{Collier:2024lys} to degenerate labels, see Appendix \ref{app-degen-rec}. It includes the same ingredients that enter the generic $5$-point amplitude and reads
\begin{align}
&\mathsf{A}_{0,6}^{(b)}(p_{1,s}) = -\sum \limits_{i=1}^5 \sum \limits_{l = \frac{1-s}{2}}^{\frac{s-1}{2}} |p_i-lb^{-1}|\mathsf{A}_{0,5}^{(b)}(p_i-lb^{-1}) + \nonumber \\
& + \frac{s-b^2}{2b} \left[\sum \limits_{i=1}^5 \sum \limits_{l = \frac{1-s}{2}}^{\frac{s-1}{2}}\mathsf{A}_{0,5}^{(b)}(p_i-lb^{-1}) - 2s \mathsf{A}_{0,5}^{(b)} + 4b S_2 C_5\left(t_1 - \sum \limits_{i=1}^5 p_i^2\right) \right] + \nonumber\\
&+2b^3 C_5 \left(-\frac{2S_3}{b^3} \sum p_i^2 + \frac{10}{3}t_1 \frac{S_3}{b^3}  - \frac{S_5}{b^5} \right) + \sum \limits_{ab|cde} \left(-\frac{2S_1}{b} \mathsf{V}_{0,4}(ip_{cde},0) + \frac{2S_3}{b^3} \right) \text{1E}^{(1)}_{ab|cde} -\nonumber \\
&- \frac{4S_1}{b} \left(\sum \limits_{ab|cde} \text{1E}^{(2)}_{cde} + \sum \limits_{ab|c|de} \text{2E}_{ab|c} \right), \label{deg6pf}
\end{align}
where
$S_k = \sum \limits_{l=\frac{1-s}{2}}^{\frac{s-1}{2}}|l|^k
$, $t_1 = \frac{b^2 + b^{-2}}{4}$ and $C_5, \text{1E}^{(k)}, \text{2E}$ were introduced in \eqref{5pf-ingredients}.

The first two contributions are explained by usual codimension-1 terms in HEM. As a check that HEM logic still applies here, let us compare codimension-2 contributions, given by the last two lines. The remainder in the curvature term can be checked similarly.

Generalizing the HEM argument we expect that the last two lines are reproduced by summing over all contributions of the form
\begin{equation}
\sum \limits_{k=\frac{1-s}{2}}^{\frac{s-1}{2}} |k|^3\int \limits_0^1 du\,u\,(1-u) \mathsf{A}_{0,3}(p_a,p_b,(u-1)\frac{k}{b})\mathsf{A}_{0,4}(p_c,p_d,p_e,u \frac{k}{b}).
\end{equation}
We expand the two factors as
\begin{equation}
\begin{cases}
 \mathsf{A}_{0,3} = \sum \limits_{m \geq 1} a^3_{0,m} \sin (2\pi m k (u-1)) = \sum \limits_{m \geq 1} a^3_{0,m} \sin (2\pi m k u)  \\
  \mathsf{A}_{0,4} = \sum \limits_{n \geq 1}( a^4_{0,n} +a^4_{1,n} (\frac{uk}{b})^2) \sin (2\pi n k u)  \\
\end{cases}, \label{a03-a04-decomp}
\end{equation}
\begin{equation}
\begin{cases}
a^3_{0,m} = 2b(-1)^m \frac{\mathcal{N}_{ab}(m)}{\sin \pi m b^2}    \\
a^4_{0,n} = 2b^2 \mathsf{V}_{0,4}(ip_{cde},0) \frac{\mathcal{N}_{cde}(m)}{\sin^2 \pi m b^2} + \frac{1}{\pi^2}\sum_{m_2=1}^\infty (-1)^{n+m_2}\frac{\mathcal{N}_{c}(n) \mathcal{N}_{de}(m_2)}{\sin(\pi n b^2)\sin(\pi m_2 b^2)} G_1(n,m_2) +\text{2 perms}  \\
a^4_{1,n} = -2b^2\frac{\mathcal{N}_{cde}(m)}{\sin^2 \pi m b^2}
\end{cases}.
\end{equation}
This leads to the integrals
\begin{multline}
\int \limits_0^1 du\,u\,(1-u) \sin 2\pi m ku \sin 2\pi nku = \frac{\delta_{mn}}{12} - \frac{G_1(m,n)}{4\pi^2 k^2},    \\
\int \limits_0^1 du\,u^3\,(1-u) \sin 2\pi m ku \sin 2\pi nku = \int \limits_0^1 du\,u\,(1-u)^3 \sin 2\pi m ku \sin 2\pi nku  = \\= \frac{\delta_{mn}}{40} - \frac{G_1(m,n)}{8\pi^2 k^2}  + \frac{3 G_2(m,n)}{4\pi^4 k^4} .
\end{multline}
The answer is then
\begin{align*}
\sum \limits_{m,n \geq 1} a_{0,m}^3 \left[ a_{0,n}^4 \left(\frac{S_3\delta_{mn}}{12} - \frac{S_1G_1(m,n)}{4\pi^2}  \right) + \frac{1}{b^2} a_{1,n}^4 \left(\frac{S_5 \delta_{m,n}}{40} - \frac{S_3 G_1(m,n)}{8\pi^2} + \frac{3S_1 G_2(m,n)}{4\pi^4} \right) \right].
\end{align*}
After summation 5 terms in the line above bring the following:
\begin{enumerate}
    \item first --- term proportional to $C_5$ and $\text{1E}^{(1)}_{abx}$, where $x=c,d,e$;
    \item second --- term proportional to $\text{1E}^{(1)}_{cde}$ and $\text{2E}_{abx}$;
    \item third --- proportional to $C_5$;
    \item fourth --- proportional to $\text{1E}^{(1)}_{cde}$;
    \item fifth --- proportional to $\text{1E}^{(2)}_{cde}$.
\end{enumerate}
All the terms are of the form appearing in the remainder of \eqref{deg6pf}; no new structures appear. Collecting everything together, one gets
\begin{align*}
&\frac{S_3}{12}\Big[4b^3V_{0,4}(p_{cde},0) C_5+2\sum_x\text{1E}^{(1)}_{abx}\Big]-S_1b^2\Big[V_{0,4}(p_{cde},0) \text{1E}^{(1)}_{cde}+\sum_x\text{2E}_{ab|x}\Big]\\
&\quad-\frac{S_5}{20}\,2bC_5+\frac{S_3}{4\pi^2}\,2\pi^2\text{1E}^{(1)}_{cde}-\frac{3S_1}{2\pi^4}\cdot\frac{4\pi^4b^2}3\text{1E}^{(2)}_{cde},
\end{align*}
which reproduces the last two lines of \eqref{deg6pf} (up to a factor $\frac{2}{b^3}$) after summing over all $10$ splits $ab|cde$ .

\section{Check of boundary period for 1GR-5T} \label{app-boundperiod}
Consider for example the period over $\Delta_{126}$, where $6$ is the position of GR operator. The 1GR-5T amplitude is the first one where the edge correction matters, but terms with $\psi$-classes in the edge correction still do not contribute. The reason is that the associated graphs $\hat{\Gamma}$ have moduli space dimension $1=\text{dim\,}\overline{\mathcal{M}}_{0,6}-2$, so we only need to pushforward a degree zero class to have a nontrivial intersection number with some boundary divisor. 

 The figure illustrates all types of graphs that contribute.

\begin{figure}[!ht]
\centering
\begin{tikzpicture}[baseline=0]
  \node[vert] (v) at (0,0) {};
  \leg{lX}{v}{155}{$2$} \leg{lX}{v}{210}{$1$} \leg{l6}{v}{90}{$6$}
  \leg{lo}{v}{35}{} \leg{lo}{v}{-35}{} \leg{lo}{v}{-90}{}
\end{tikzpicture}
\qquad\qquad
\begin{tikzpicture}[baseline=0]
  \node[vert] (v) at (0,0) {}; \node[vert] (w) at (150:0.8) {};
  \draw[edge] (v) -- (w);
  \leg{lX}{w}{175}{$2$} \leg{l6}{w}{90}{$6$}
  \leg{lX}{v}{210}{$1$}
  \leg{lo}{v}{35}{} \leg{lo}{v}{-35}{} \leg{lo}{v}{-90}{}
\end{tikzpicture}
\quad
\begin{tikzpicture}[baseline=0]
  \node[vert] (a) at (0,0) {}; \node[vert] (b) at (1.2,0) {};
  \draw[edge] (a) -- (b);
  \leg{lX}{a}{145}{$2$} \leg{lX}{a}{215}{$1$} \leg{l6}{a}{90}{$6$}
  \threeright{b}
\end{tikzpicture}
\quad
\begin{tikzpicture}[baseline=0]
  \node[vert] (a) at (0,0) {}; \node[vert] (b) at (1.2,0) {};
  \draw[edge] (a) -- (b);
  \leg{lX}{a}{145}{$2$} \leg{lX}{a}{215}{$1$}
  \leg{l6}{b}{100}{$6$} \threeright{b}
\end{tikzpicture}
\quad
\begin{tikzpicture}[baseline=0]
  \node[vert] (a) at (0,0) {}; \node[vert] (b) at (1.2,0) {}; \node[vert] (w) at (135:0.7) {};
  \draw[edge] (a) -- (b); \draw[edge] (a) -- (w);
  \leg{lX}{w}{175}{$2$} \leg{l6}{w}{90}{$6$}
  \leg{lX}{a}{215}{$1$}
  \threeright{b}
\end{tikzpicture}
\quad
\begin{tikzpicture}[baseline=0]
  \node[vert] (a) at (0,0) {}; \node[vert] (m) at (0.9,0) {}; \node[vert] (b) at (1.8,0) {};
  \draw[edge] (a) -- (m) -- (b);
  \leg{lX}{a}{145}{$2$} \leg{lX}{a}{215}{$1$}
  \leg{l6}{m}{90}{$6$}
  \threeright{b}
\end{tikzpicture}
\quad
\begin{tikzpicture}[baseline=0]
  \node[vert] (a) at (0,0) {}; \node[vert] (b) at (1.2,0) {};
  \draw[edge] (a) -- (b);
  \leg{lX}{a}{135}{$2$} \leg{lX}{a}{180}{$1$} \leg{l6}{a}{90}{$6$} \leg{lo}{a}{235}{}
  \tworight{b}
\end{tikzpicture}
\qquad\qquad
\begin{tikzpicture}[baseline=0]
  \node[vert] (a) at (0,0) {}; \node[vert] (b) at (1.2,0) {}; \node[vert] (w) at (125:0.75) {};
  \draw[edge] (a) -- (b); \draw[edge] (a) -- (w);
  \leg{lX}{w}{170}{$2$} \leg{l6}{w}{90}{$6$}
  \leg{lX}{a}{180}{$1$} \leg{lo}{a}{235}{}
  \tworight{b}
\end{tikzpicture}
\quad
\begin{tikzpicture}[baseline=0]
  \node[vert] (a) at (0,0) {}; \node[vert] (m) at (1.0,0) {}; \node[vert] (b) at (2.0,0) {};
  \draw[edge] (a) -- (m) -- (b);
  \leg{lX}{a}{145}{$2$} \leg{lX}{a}{215}{$1$} \leg{l6}{a}{95}{$6$}
  \leg{lo}{m}{90}{}
  \tworight{b}
\end{tikzpicture}
\qquad\qquad
\begin{tikzpicture}[baseline=0]
  \node[vert] (a) at (0,0) {}; \node[vert] (m) at (1.0,0) {}; \node[vert] (b) at (2.0,0) {};
  \draw[edge] (a) -- (m) -- (b);
  \leg{lX}{a}{145}{$2$} \leg{lX}{a}{215}{$1$}
  \leg{lo}{m}{90}{} \leg{l6}{m}{-90}{$6$}
  \tworight{b}
\end{tikzpicture}
\end{figure}

For each graph, as mentioned in the main text, we either already have an edge separating $\lbrace 1,2,6\rbrace$ from other legs (``excess intersection'') or an additional pullback to moduli space, associated to refined graph $\hat{\Gamma}$, is needed. For each graph we will only mention which option is realized and provide the final result of the calculation
\begin{enumerate}
    \item pullback to $\Delta_{126} \cong \overline{\mathcal{M}}_{0,4} \times \overline{\mathcal{M}}_{0,4}$; result
\begin{equation}
2b^3 C_5 s (t_1 - p_1^2 - p_2^2 -\frac{\left(b^4+s^2\right)}{4 b^2}) (t_1 - \sum \limits_{i=3}^5 p_i^2 ).
\end{equation}
\item pullback to the graph with two edges of type $26|1|345$. For the graph in the figure we get
\begin{equation}
2b^3 C_5(s p_2^2 + \frac{S_2}{b^2}) (t_1 - \sum \limits_{i=3}^5 p_i^2 ),
\end{equation}
which we sum with an analogous term with $p_2 \to p_1$. 

\item excess intersection;
\begin{equation}
-s \left[(t_1  -p_1^2  - p_2^2  -\frac{(b^4+s^2)}{4b^2}  )  + (t_1 -  \sum \limits_{i=3}^5 p_i^2 )\right] \text{1E}_{345}^{(1)}-2s\,\text{1E}^{(2)}_{345}.
\end{equation}

\item pullback to the graph of type $12|6|345$; $\overline{\mathcal{M}}_{\hat{\Gamma}} \cong \overline{\mathcal{M}}_{0,4}$. Result
\begin{equation}
    s(t_1 - \sum \limits_{i=3}^5 p_i^2) \text{1E}_{345}^{(1)}.
\end{equation}
\item excess intersection. From two placements of leg $6$ in total we get
$$
-  (s p_2^2 + \frac{S_2}{b^2}) \cdot \text{1E}_{345}^{(1)} -  (s p_1^2 + \frac{S_2}{b^2}) \cdot \text{1E}_{345}^{(1)}.
$$
\item excess intersection again; the result
\begin{equation}
2s\, \text{1E}_{345}^{(2)} - \frac{2S_2}{b^2}\text{1E}_{345}^{(1)} + 2b^3 C_5 \frac{S_4}{6b^4}.
\end{equation}
\item pullback to the graph of the type $126|e|cd$, where $e$ is the third leg on left;
\begin{equation}
s(t_1 - \sum \limits_{i=1,2} p_i^2 -\frac{\left(b^4+s^2\right)}{4 b^2}) \text{1E}_{12e}^{(1)}.
\end{equation}

\item pullback to the graph of the type $26|1|e|cd$; sum of answers for leg $6$ attached to $1,2$
\begin{equation}
(sp_1^2 + sp_2^2 + \frac{2S_2}{b^2}) \sum \limits_{e=3,4,5}\text{1E}_{12e}^{(1)}.
\end{equation}
\item the final two contributions cancel each other --- first one corresponds to excess intersection, which brings the $(-1)$ sign, the second can be pulled back to $12|6|e|cd$ graph. They cancel for each choice of the middle leg $e$.
\end{enumerate} 
Collecting all of this together, we obtain the period
\begin{equation}
 -\frac{S_2}{b^2}\left[2b^3 C_5 V_{0,4}(p_{345},0)+ \sum \limits_{c} \text{1E}^{(1)}_{12c} + \text{1E}^{(1)}_{345}  \right] + \frac{S_4}{6b^4}\cdot 2b^3 C_5. \label{bound-per}
\end{equation}
Now we compute the expected answer. Up to normalization it is given by
\begin{equation}
\sum \limits_{k=\frac{1-s}{2}}^{\frac{s-1}{2}} k^2\int \limits_0^1 du\,(1-u) \mathsf{A}_{0,3}(p_1,p_2,(u-1)\frac{k}{b})\mathsf{A}_{0,4}(p_c,p_d,p_e,u \frac{k}{b}).
\end{equation}
Using decomposition \eqref{a03-a04-decomp} of three- and four-point amplitudes as a function of $u$-dependent momentum and the integrals
\begin{multline}
\int \limits_0^1 du\,(1-u) \sin 2\pi m k u \sin 2\pi n ku = \frac{\delta_{mn}}{4}, \\
\int \limits_0^1 du\,(1-u) u^2 \sin 2\pi m k u \sin 2\pi n ku =\frac{\delta_{mn}}{24}  - \frac{G_1(m,n)}{8k^2 \pi^2},
\end{multline}
we get
\begin{align*}
&\frac{S_4}{24b^2}(-4b^3)C_5 + \frac{S_2}{4} 4b^3 C_5 \mathsf{V}_{0,4}(ip_{345},0) + \\
&+ \sum \limits_c \frac{S_2}{4\pi^2} 2b \sum_{m_2=1}^\infty (-1)^{m_2}\frac{\mathcal{N}_{12c}(m) \mathcal{N}_{de}(m_2)}{\sin^2(\pi m b^2)\sin(\pi m_2 b^2)} G_1(n,m_2) + \\
&+4b^3\frac{S_2}{8\pi^2 b^2} \sum \limits_{m,n} G_1(m,n) (-1)^{m}\frac{\mathcal{N}_{12}(m) \mathcal{N}_{345}(n)}{\sin^2(\pi n b^2)\sin(\pi m b^2)}.
\end{align*}
This is the same as \eqref{bound-per} up to coefficient.

 \bibliographystyle{JHEP}
\bibliography{bib}
\end{document}